\documentclass[11pt]{article}%
\usepackage{amssymb,amsmath,xcolor,graphicx,xspace,colortbl, rotating}
\usepackage{amsfonts}
\usepackage{amsthm}
\usepackage{caption}
\usepackage[mag=1000,portrait]{geometry}
\usepackage{indentfirst}
\usepackage[onehalfspacing]{setspace}
\usepackage{threeparttable}
\usepackage{subcaption}
\usepackage{booktabs}
\usepackage{amsmath}
\usepackage{amssymb}
\usepackage{graphicx}%
\usepackage[justification=centering]{caption}
\usepackage{ulem}
\usepackage{todonotes}
\providecommand{\U}[1]{\protect\rule{.1in}{.1in}}
\DeclareGraphicsExtensions{.pdf,.eps,.ps,.png,.jpg,.jpeg}
\providecommand{\U}[1]{\protect\rule{.1in}{.1in}}
\newtheorem{theorem}{Theorem}

\newtheorem{assumption}{Assumption}

\newtheorem{lemma}{Lemma}

\DeclareMathOperator*{\argmin}{arg\,min}

\begin{document}

\author{$%
	\begin{array}
	[c]{ccc}%
	\text{Jungbin Hwang}\thanks{Email: jungbin.hwang@uconn.edu}& \hspace{0.2in} & \text{Byunghoon Kang}\thanks{Kang acknowledges the research support from Lancaster University Management School (Research Pump-Prime Grants ECA6396), Email: b.kang1@lancaster.ac.uk}\\
	\text{University of Connecticut} & \hspace{0.2in} & \text{Lancaster University}\\
	& &\\
	\multicolumn{3}{c}{\text{Seojeong Lee}\thanks{Lee acknowledges that this research was supported under the
			Australian Research Council Discovery Early Career Reserach Award (DECRA)
			funding scheme (project number DE170100787), Email: jay.lee@unsw.edu.au}}\\
	\multicolumn{3}{c}{\text{University of New South Wales}}
	\end{array}
	\medskip$}
\title{A Doubly Corrected Robust Variance Estimator for Linear GMM}
\date{\today \footnote{We thank the Co-Editor Xiaohong Chen, an anonymous Associate Editor, and an anonymous referee for valuable comments on a previous version. We also thank Frank Windmeijer, Atsushi Inoue, Bruce Hansen, Frank Kleibergen, Whitney Newey, Richard Smith, Yixiao Sun and participants at ANZESG2019, ESAM2019, SETA2019, NASMES2019, IPDC2019,  KER 2019, ESEM 2019 and CFE 2019 for helpful comments and suggestions.}}
\maketitle

\begin{abstract}
We propose a new finite sample corrected variance estimator for the linear generalized method of moments (GMM) including the one-step, two-step, and iterated estimators. Our formula additionally corrects for the over-identification bias in variance estimation on top of the commonly used finite sample correction of Windmeijer (2005) which corrects for the bias from estimating the efficient weight matrix, so is\textit{ doubly} corrected. An important feature of the proposed double correction is that it automatically provides robustness to misspecification of the moment condition. In contrast, the conventional variance estimator and the Windmeijer correction are inconsistent under misspecification. That is, the proposed double correction formula provides a convenient way to obtain improved inference under correct specification and robustness against misspecification at the same time.  
\end{abstract}

\section{Introduction}

\indent
The generalized method of moments (GMM) estimators (Hansen, 1982) are widely used in economics. Among the class of GMM estimators, the efficient GMM has the smallest asymptotic variance which can be obtained via a two-step procedure. However, researchers have found that the standard error of the two-step efficient GMM is often severely downward biased. To solve this problem, Windmeijer (2005) proposed a finite sample bias-corrected standard error formula for the two-step linear GMM. Specifically, his formula corrects for the bias arising from using the efficient weight matrix being evaluated at an estimate, rather than the true value. The correction formula (the Windmeijer correction, hereinafter) has been routinely used in practice.\footnote{More than 5,200 citations according to Google Scholar on May 26, 2020.}

However, the Windmeijer correction does not take into account the \textit{over-identification bias}, which is another important source of bias in the GMM standard error. The over-identification bias arises from the fact that the over-identified sample moment condition is nonzero in general while it converges in probability to zero under correct specification. 

We propose a new finite sample correction which takes into account the over-identification bias for the variance of the linear one-step, two-step, and iterated GMM estimators. For one-step GMM such as the two-stage least squares (2SLS) estimators, the proposed finite sample correction is new as the Windmeijer correction does not cover the one-step GMM. For two-step and iterated GMM, the proposed correction improves upon the Windmeijer correction by additionally correcting for the over-identification bias. Thus, we \textit{doubly} correct the finite sample bias of the linear GMM variance estimator. 

The order of our double correction terms equals the order of the sample moment condition. Under correct specification or local misspecification (where the moment condition is modeled as a drifting sequence within a $n^{-1/2}$-neighborhood), these terms are $O_{p}(n^{-1/2})$ so that the double correction is a finite-sample correction for the variance. We provide a stochastic expansion of the GMM estimators under local misspecification in the appendix which shows that the double correction estimates the (co)variance of higher-order terms which increase with the over-identification bias.

Under (global) misspecification, however, the double correction terms no longer degenerate to zero asymptotically because the stochastic order of the sample moment condition becomes $O_{p}(1)$. The conventional variance estimator and the Windmeijer correction omit these $O_{p}(1)$ terms under misspecification. This implies that the conventional variance estimator and the Windmeijer correction are inconsistent, while our doubly corrected variance estimator is consistent regardless of whether the moment condition model is (locally or globally) misspecified or not.

Since the doubly corrected variance estimators are robust to misspecification, it is not surprising that the formulas coincide with the misspecification-robust variance estimator in Lee (2014) for the one-step and two-step GMM and Hansen and Lee (2019) for the iterated GMM. Indeed the simulation results reported in those papers show that the misspecification-robust variance estimator often performs better than the conventional sandwich variance estimator under correct specification. This paper provides an answer to this seemingly puzzling result by taking an alternative path to obtain the misspecification-robust variance estimator formula. Our approach provides new insight into the misspecification-robust formula as a finite-sample correction closely related to the well-known Windmeijer (2005) correction. We show that the misspecification-robust variance estimators of Lee (2014) and Hansen and Lee (2019) provide the same order of finite-sample correction with the Windmeijer (2005) correction under correct specification and local misspecification. To the best of our knowledge, this paper is the first to show the equivalence between the finite-sample corrected variance formula and the robust variance formula in misspecified GMM.

From a practical point of view, this implies that accurate inference under correct specification and robust inference under misspecification can be achieved simultaneously, without knowing whether the moment condition is correctly specified or not. Moreover, it can be easily implemented to obtain more accurate $t$ tests and confidence intervals (smaller errors in the size and the coverage) by bootstrapping the $t$ statistic studentized by the doubly corrected variance estimator. Lee (2014) shows that this bootstrap procedure is robust to misspecification and does not require an ad hoc correction in the bootstrap sample called recentering.

The finite sample correction of the proposed formula and the Windmeijer formula work for linear models. For nonlinear models, the order of the remainder term is the same as the correction terms, so that the corrections do not necessarily provide improvements under correct specification.

Robust inference with possibly misspecified moment condition models has gained considerable attention in the literature. For linear instrumental variable (IV) models, Maasoumi and Phillips (1982) investigate the limiting distribution of inconsistent IV estimators. Guggenberger (2012) studies the behavior of the weak instrument robust tests under local misspecification. Kang (2018) derives higher-order expansions of IV estimators allowing for local violation of the instrument validity condition. Lee (2018) shows that the moment condition is misspecified under treatment effect heterogeneity and proposes a robust variance estimator for 2SLS.

For general moment condition models, Hall and Inoue (2003) derive the asymptotic distribution of GMM under misspecification. Schennach (2007) proposes an alternative GEL-type estimator robust to global misspecification. Ai and Chen (2007) investigate the asymptotic properties of the sieve minimum distance estimator under misspecified conditional moment restrictions model. Otsu (2011) analyses moderate deviation behaviors of GMM.  Kitamura, Otsu, and Evdokimov (2013) propose an estimator that achieves optimal minimax robust properties under local misspecification. Lee (2014, 2016) propose a robust nonparametric bootstrap procedure for GMM and GEL estimators. Hansen and Lee (2019) provide robust inference theory for the iterated GMM. Rotemberg (1983) and Andrews (2019) characterize the estimands of the linear GMM under misspecification. Andrews, Gentzkow, and Shapiro (2017) propose to measure the effect of model misspecification on the sensitivity of parameter estimates for the minimum distance estimators. Bonhomme and Weidner (2018) and Armstrong and Koles\'{a}r (2019) consider minimax and GMM inference under possible misspecification, respectively.

Finite sample properties of GMM estimators, including the iterated and the continuously updating (CU) GMM are investigated by Hansen, Heaton, and Yaron (1996). Bond and Windmeijer (2005) provide simulation evidence on the finite sample performance of the asymptotic and bootstrap tests based on GMM estimators. Hwang (2020) develops fixed-cluster asymptotics and finite-sample corrected variance formula for cross-sectionally dependent data. Hwang and Sun (2018) employ fixed-smoothing asymptotics to provide a more accurate comparison between the one-step and two-step GMM procedures for time-series observations.

Our doubly corrected robust variance estimators are generally different than the many instruments and many weak instruments robust variance estimators for IV, GMM, and GEL estimators proposed by Bekker (1994), Han and Phillips (2006), Newey and Windmeijer (2009), and Evdokimov and Koles\'{a}r (2018). Since our double correction formula does not use the many (weak) instruments asymptotics, it is not robust under such sequences.  

The remainder of the paper is organized as follows. Section \ref{Section: Windmeijer_Correction} reviews the Windmeijer correction. Section \ref{Section_Double_Correction} proposes the doubly corrected variance estimator. Section \ref{Misspecification} shows that the doubly corrected variance estimator is misspecification-robust. Section \ref{Section_iteragedGMM} discusses the iterated GMM and the CU GMM. Section \ref{Section_Examples} derives the double correction formula for cross-sectional IV and the difference GMM. Finally, Section \ref{Section: Simulation} provides extensive simulation results comparing the double correction and other variance estimators. All the proofs are collected in Appendix A. In Appendix B, we derive the stochastic expansion of the one-step and two-step GMM estimators under local misspecification and show that the double correction estimates the (co)variance of some higher-order terms.

\section{Finite Sample Correction of Windmeijer (2005)\label{Section: Windmeijer_Correction}}

Suppose that we observe a sequence of i.i.d. random vectors $X_{i}\in \mathbb{R}^{d_{x}}$ for $i=1,...,n$. Let $g(X_{i},\theta)$ be a $q\times1$ moment function where $\theta$ is a $k\times1$ parameter vector. We assume $q>k$ so that the model is over-identified and $g(X_{i},\theta)$ is linear in parameter. When the model is just-identified ($q=k$) the correction terms are zero and the analysis becomes trivial. The moment condition model is correctly specified if 
\begin{equation}
E[g(X_{i},\theta_{0})]=0
\label{correct_spec}
\end{equation}
for a unique $\theta_{0}$. Assume $E[\|g(X_{i},\theta_{0})\|^{2}]<\infty$ so that
\begin{equation}
\frac{1}{n}\sum_{i=1}^{n}g(X_{i},\theta_{0}) \equiv g_{n}(\theta_{0})= O_{p}(n^{-1/2}).
\label{sample_mean_moment}
\end{equation}
Thus, the sample moment condition converges in probability to zero at the rate of $n^{-1/2}$ under correct specification \eqref{correct_spec}. This will be used in determining the order of higher-order terms in Sections 2 and 3. 


The one-step GMM estimator is defined as
\begin{equation}
\hat{\theta}_{1}=\underset{\theta\in\Theta}{\arg\min}~g_{n}(\theta)^{\prime
}W_{n}^{-1}g_{n}(\theta),
\end{equation}
where $W_{n}$ is a $q\times q$ positive definite weight matrix which takes the form of $n^{-1}\sum_{i=1}^{n}W(X_{i})$ and $W(X_{i})$ does not depend on any unknown parameter. Common choices of $W(X_{i})$ are the identity matrix and $Z_{i}Z_{i}^{\prime}$ where $Z_{i}$ is the instrument vector in IV regressions. Let $W=EW_{n}$, a positive definite matrix of constants.

The two-step efficient GMM estimator using $\hat{\theta}_{1}$ as a preliminary (initial) estimator is defined as
\begin{equation}
\hat{\theta}_{2}=\arg\min_{\theta\in\Theta}~g_{n}(\theta)
^{\prime}[\Omega_{n}(\hat{\theta}_{1})]^{-1}g_{n}(\theta),
\label{GMM_two_step_uncentered}%
\end{equation}
where 
\begin{equation*}
\Omega_{n}(\theta)=\frac{1}{n}\sum_{i=1}^{n}g(X_{i},\theta)g(X_{i},\theta)^{\prime}.
\end{equation*}
Define $\Omega =\Omega(\theta_{0})$ where $\Omega(\theta) = E\Omega_{n}(\theta)$. Since $\Omega_{n}(\hat{\theta}_{1})$ is consistent for the asymptotic variance of the moment function the two-step GMM is efficient. 

We also define an infeasible two-step GMM estimator $\tilde{\theta}_{2}$ using $[\Omega_{n}(\theta_{0})]^{-1}$ as the weight matrix:
\begin{equation}
\tilde{\theta}_{2}=\arg\min_{\theta\in\Theta}~g_{n}(\theta)^{\prime}%
[\Omega_{n}(\theta_{0})]^{-1}g_{n}(\theta).
\end{equation}
Investigating the limiting behavior of $\sqrt{n}(\tilde{\theta}_{2}-\theta_{0})$ will help us understand the higher-order behavior of the
feasible two-step estimator $\sqrt{n}(\hat{\theta}_{2}-\theta_{0})$.

Let $G(X_{i})=\partial g(X_{i},\theta)/\partial\theta^{\prime}$. Note that it does not
depend on $\theta$ due to linearity. Define
$G_{n}=n^{-1}\sum_{i=1}^{n}G(X_{i})$ and $G=EG_{n}$, which is assumed full column-rank. By the first-order Taylor expansion, the first-order condition (FOC) of the (feasible) two-step GMM can be written as
\begin{align}
\label{twofoc}0  &  =G_{n}^{\prime}[\Omega_{n}(\hat{\theta}_{1})]^{-1}g_{n}(\hat{\theta
}_{2})\\
\notag &  =G_{n}^{\prime}[\Omega_{n}(\hat{\theta}_{1})]^{-1}\left[  g_{n}%
(\theta_{0})+G_{n}(\hat{\theta}_{2}-\theta_{0})\right]  ,
\end{align}
and so we have
\begin{equation}
\sqrt{n}(\hat{\theta}_{2}-\theta_{0})=-\left\{  G_{n}^{\prime}%
[\Omega_{n}(\hat{\theta}_{1})]^{-1}G_{n}\right\}  ^{-1}G_{n}^{\prime
}[\Omega_{n}(\hat{\theta}_{1})]^{-1}\sqrt{n}g_{n}(\theta_{0}).
\label{Eq_Mean_Value_Feasible_Two_Step}%
\end{equation}
Using a similar expansion$,$ we can get
\begin{equation}
\sqrt{n}(\tilde{\theta}_{2}-\theta_{0})=-\left\{  G_{n}^{\prime
}[\Omega_{n}(\theta_{0})]^{-1}G_{n}\right\}  ^{-1}G_{n}^{\prime}%
[\Omega_{n}(\theta_{0})]^{-1}\sqrt{n}g_{n}(\theta_{0}),
\label{Eq_Mean_Value_Infeasible_Two_Step}%
\end{equation}
for the infeasible two-step GMM and
\begin{equation}
\sqrt{n}(\hat{\theta}_{1}-\theta_{0})=-(G_{n}^{\prime}W_{n}^{-1}%
G_{n})^{-1}G_{n}^{\prime}W_{n}^{-1}\sqrt{n}g_{n}(\theta_{0})
\label{Eq_Mean_Value_First_Step}%
\end{equation}
for the one-step GMM.

Asymptotically \eqref{Eq_Mean_Value_Feasible_Two_Step} and \eqref{Eq_Mean_Value_Infeasible_Two_Step} have the same limiting distribution so that using $\Omega_{n}(\hat{\theta}_{1})$ instead of $\Omega_{n}(\theta_{0})$ does not affect the first-order asymptotic analysis. However, by expanding $\Omega_{n}(\hat{\theta}_{1})$ around $\theta_{0}$ and using \eqref{Eq_Mean_Value_First_Step}, Windmeijer (2005) shows that the extra finite sample variations caused by higher-order terms can be estimated and the accuracy of the variance estimator can be improved for linear moment condition models.

To see this, we use the first-order Taylor expansion of $\Omega_{n}(\hat{\theta}_{1})$ in the right-hand side (RHS) of \eqref{Eq_Mean_Value_Feasible_Two_Step} around $\theta_{0}$:
\begin{align}
\notag \sqrt{n}(\hat{\theta}_{2}-\theta_{0}) &  =-\left\{  G_{n}^{\prime}%
[\Omega_{n}(\theta_{0})]^{-1}G_{n}\right\}  ^{-1}G_{n}^{\prime}[\Omega_{n}(\theta_{0})]^{-1}\sqrt{n}g_{n}(\theta_{0})+D_{n}  \sqrt{n}(\hat{\theta}_{1}-\theta_{0})+R_{n}\\
&  =\sqrt{n}(\tilde{\theta}_{2}-\theta_{0})+\underset{=O_{p}(n^{-1/2})}{\underbrace{D_{n}  \sqrt{n}(\hat{\theta}_{1}-\theta_{0})}}+R_{n},
\label{Eq_feasible2_Expansion}
\end{align}
where%
\begin{align*}
D_{n} &= F_{1n} + F_{2n},\\
F_{1n} &  = - \left.  \frac{\partial\left\{  G_{n}^{\prime}%
[\Omega_{n}(\theta)]^{-1}G_{n}\right\}  ^{-1}}{\partial\theta^{\prime}%
}\right\vert _{\theta=\theta_{0}}G_{n}^{\prime}[\Omega_{n}(\theta
_{0})]^{-1}g_{n}(\theta_{0}),\\
F_{2n} &  =- \left\{  G_{n}^{\prime}[\Omega_{n}(\theta_{0}%
)]^{-1}G_{n}\right\}  ^{-1}\left.  \frac{\partial \{G_{n}^{\prime}%
[\Omega_{n}(\theta)]^{-1}g_{n}(\theta_{0})\}}{\partial\theta^{\prime}%
}\right\vert _{\theta=\theta_{0}}%
\end{align*}
are $k\times k$ matrices and $R_{n}$ is the remainder term. Since $g_{n}(\theta_{0})=O_{p}(n^{-1/2})$ both $F_{1n}$ and $F_{2n}$ are $O_{p}(n^{-1/2})$. Thus, the second term in the RHS of \eqref{Eq_feasible2_Expansion} is of order $O_{p}(n^{-1/2})$ assuming that $\sqrt{n} (\hat{\theta}_{1}-\theta_{0}) = O_{p}(1)$. The remainder term $R_{n}$ is of order $O_{p}(n^{-1})$ because of the linearity of the moment function provided that the higher moments of $g(X_{i},\theta_{0})$ and $G(X_{i})$ exist (the formal justification is given in the proof of Theorem 1). Thus, by taking into account for the variation caused by the $O_{p}(n^{-1/2})$ term, the finite sample variance of $\sqrt{n}(\hat{\theta}_{2}-\theta_{0})$ can be more accurately approximated. Note that the expansion \eqref{Eq_feasible2_Expansion} only holds for linear moment condition models. 

The Windmeijer correction of the variance of $\sqrt{n}(\hat{\theta}_{2}-\theta_{0})$ is obtained by 
\begin{equation}
\widehat{V}_{w}(\hat{\theta}_{2}) =\widetilde{V}(\hat{\theta}_{2}) 
+\widehat{D}_{n}\widetilde{V}(\hat{\theta}_{2}) +\widetilde{V}(\hat{\theta}_{2})\widehat{D}_{n}^{\prime}+\widehat{D}_{n}\widetilde{V}(\hat{\theta}_{1})\widehat{D}_{n}^{\prime},
\label{Variance_Windmeijer}
\end{equation}
where $D[.,j]$ denotes the $j$th column of $D$, $\theta_{[j]}$ denotes the $j$th element of $\theta$, and
\begin{align*}
\widetilde{V}(\hat{\theta}_{1})&=\left(  G_{n}^{\prime}W_{n}%
^{-1}G_{n}\right)  ^{-1}\left(  G_{n}^{\prime}W_{n}^{-1}\Omega_{n}%
(\hat{\theta}_{1})W_{n}^{-1}G_{n}\right)  \left(  G_{n}^{\prime}W_{n}%
^{-1}G_{n}\right)  ^{-1},\\
\widetilde{V}(\hat{\theta}_{2})&=\left\{  G_{n}^{\prime}[\Omega_{n}(\hat{\theta}_{1})]^{-1}G_{n}\right\}  ^{-1},\\
\widehat{D}_{n}[.,j]& =\left\{  G_{n}^{\prime}[\Omega_{n}%
(\hat{\theta}_{1})]^{-1}G_{n}\right\}  ^{-1}G_{n}^{\prime}\left\{
[\Omega_{n}(\hat{\theta}_{1})]^{-1}\left.  \frac{\partial\Omega_{n}%
	(\theta)}{\partial\theta_{[j]}}\right\vert _{\theta=\hat{\theta}_{1}}%
[\Omega_{n}(\hat{\theta}_{1})]^{-1}\right\}  g_{n}(\hat{\theta}_{2}),\\
\frac{\partial\Omega_{n}(\theta)}{\partial\theta_{[j]}} &  =\Upsilon
_{j}(\theta)+\Upsilon_{j}^{\prime}(\theta),\nonumber\\
\Upsilon_{j}(\theta) &  =\frac{1}{n}\sum_{i=1}^{n}g(X_{i},\theta)\frac{\partial g(X_{i},\theta)}{\partial\theta_{[j]}%
}^{\prime}.\nonumber
\end{align*}
Since the estimate of $F_{1n}$ equals to zero because of the FOC, $0=G_{n}^{\prime}[\Omega_{n}(\hat{\theta}_{1})]^{-1}g_{n}(\hat{\theta}_{2})$, it does not appear in the variance estimator formula. The standard error is obtained by taking the diagonal elements of $\sqrt{\widehat{V}_{w}(\hat{\theta}_{2})/n}$.

\section{Double Correction \label{Section_Double_Correction}}

The Windmeijer correction accounts for the extra variability due to using an estimated parameter in the weight matrix. This correction is effective because $\widehat{D}_{n}\neq0$, which is due to $g_{n}(\hat{\theta}_{2})\neq0$ in finite sample. In fact, $g_{n}(\theta)\neq0$ for all $\theta$ almost surely if at least one of the moments is continuously distributed, which (trivially) implies $g_{n}(\theta_{0})\neq0$. We call this the \textit{over-identification bias}, which is non-zero for any $n$ in general for over-identified models. 

We show that the over-identification bias causes additional finite sample variability in \eqref{Eq_feasible2_Expansion}. These additional terms are not considered in the Windmeijer correction \eqref{Variance_Windmeijer}. We propose alternative variance estimators that fully incorporate the additional variations induced by the over-identification bias. These variance estimators will replace $\widetilde{V}(\hat{\theta}_{2})$ and $\widetilde{V}(\hat{\theta}_{1})$ in \eqref{Variance_Windmeijer} without affecting the order of finite sample corrections, leading to our doubly corrected variance estimator.

Assume that 
\begin{align}
\label{G_Normal} G_{n}-G =& O_{p}(n^{-1/2}),\\
\label{W_Normal} \text{vec}(W_{n}-W) =& O_{p}(n^{-1/2}),\\
\text{vec}(\Omega_{n}(\theta_{0})-\Omega) =& O_{p}(n^{-1/2}),
\label{Omega_Normal}
\end{align}
which hold under appropriate regularity conditions. Since $G'\Omega^{-1}g=0$ by the population FOC and
\begin{equation}
[\Omega_{n}(\theta_{0})]^{-1}-\Omega^{-1}=-\Omega^{-1}\left(\Omega_{n}(\theta_{0})-\Omega\right)[\Omega_{n}(\theta_{0})]^{-1},
\end{equation} 
we can write
\begin{align}
\notag & G_{n}^{\prime}[\Omega_{n}(\theta_{0})]^{-1}g_{n}(\theta_{0})\\
&= \underset{=O_{p}(n^{-1/2})}{\underbrace{G^{\prime}\Omega^{-1}g_{n}(\theta_{0})}} + \underset{=O_{p}(n^{-1})}{\underbrace{(G_{n}-G)^{\prime}\Omega^{-1}g_{n}(\theta_{0}) - G^{\prime}\Omega^{-1}\left(\Omega_{n}(\theta_{0})-\Omega\right)\Omega^{-1}g_{n}(\theta_{0})}} + O_{p}(n^{-3/2}).
\label{Mis_Expan}
\end{align}
Using \eqref{Mis_Expan}, the expansion of the infeasible two-step GMM \eqref{Eq_Mean_Value_Infeasible_Two_Step} can be written as
\begin{align}
\label{T2_Exp1}\sqrt{n}(\tilde{\theta}_{2}-\theta_{0})=&-\left\{  G_{n}^{\prime}[\Omega_{n}(\theta_{0})]^{-1}G_{n}\right\}  ^{-1}\left[G^{\prime}\Omega^{-1}\sqrt{n}g_{n}(\theta_{0})\right.\\
&\left. +\sqrt{n}(G_{n}-G)^{\prime}\Omega^{-1}g_{n}(\theta_{0}) - G^{\prime}\Omega^{-1}\sqrt{n}\left(\Omega_{n}(\theta_{0})-\Omega\right)\Omega^{-1}g_{n}(\theta_{0})\right]\\
\label{T2_Exp2}& + O_{p}(n^{-1}).
\end{align}
Similarly,
\begin{align}
\label{T1_Exp1}\sqrt{n}(\hat{\theta}_{1}-\theta_{0})=&-\left\{  G_{n}^{\prime}W_{n}^{-1}G_{n}\right\}  ^{-1}\left[G^{\prime}W^{-1}\sqrt{n}g_{n}(\theta_{0}) \right.\\
\label{T1_Exp12} &\left.+\sqrt{n}(G_{n}-G)^{\prime}W^{-1}g_{n}(\theta_{0}) - G^{\prime}W^{-1}\sqrt{n}\left(W_{n}-W\right)W^{-1}g_{n}(\theta_{0})\right]\\
\label{T1_Exp2}& + O_{p}(n^{-1}),
\end{align}
which simplifies to
\begin{equation}
\sqrt{n}(\hat{\theta}_{1}-\theta_{0})=-\left\{  G_{n}^{\prime}G_{n}\right\}  ^{-1}\left[G^{\prime}\sqrt{n}g_{n}(\theta_{0})+\sqrt{n}(G_{n}-G)^{\prime}g_{n}(\theta_{0}) \right]
\end{equation}
when $W_{n}=I$. From the above expansions we learn the followings. First, we need to consider the extra variations from $\sqrt{n}(G_{n}-G)$ and $\sqrt{n}(\Omega_{n}(\theta)-\Omega)$ (or $\sqrt{n}(W_{n}-W)$ for the one-step GMM) to account for the over-identification bias. Second, the order of the remainder term of the original expansion \eqref{Eq_feasible2_Expansion} is not changed. 

Using the expansions \eqref{T2_Exp1}-\eqref{T2_Exp2} an \eqref{T1_Exp1}-\eqref{T1_Exp2}, the expansion of the two-step GMM can be written as
\begin{align}
\label{Complete_Exp1}\sqrt{n}(\hat{\theta}_{2}-\theta_{0}) = & -\left\{  G_{n}^{\prime}[\Omega_{n}(\theta_{0})]^{-1}G_{n}\right\}  ^{-1}\left[G^{\prime}\Omega^{-1}\sqrt{n}g_{n}(\theta_{0})\right.\\
\label{Twostep_Expansion} &\left.  +\sqrt{n}(G_{n}-G)^{\prime}\Omega^{-1}g_{n}(\theta_{0})- G^{\prime}\Omega^{-1}\sqrt{n}\left(\Omega_{n}(\theta_{0})-\Omega\right)\Omega^{-1}g_{n}(\theta_{0})\right]\\
\label{Onestep_influence}&-D_{n}\left\{  G_{n}^{\prime}W_{n}^{-1}G_{n}\right\}  ^{-1}\left[G^{\prime}W^{-1}\sqrt{n}g_{n}(\theta_{0})\right.\\
\label{Onestep_Expansion}&\left. +\sqrt{n}(G_{n}-G)^{\prime}W^{-1}g_{n}(\theta_{0}) - G^{\prime}W^{-1}\sqrt{n}\left(W_{n}-W\right)W^{-1}g_{n}(\theta_{0})\right]\\
\label{Complete_Exp2}&+ O_{p}(n^{-1}).
\end{align}
Note that \eqref{Complete_Exp2} is the sum of \eqref{T2_Exp2}, \eqref{T1_Exp2}, and $R_{n}$ in \eqref{Eq_feasible2_Expansion}. The first two terms are $O_{p}(n^{-1})$ under \eqref{G_Normal}-\eqref{Omega_Normal}. 

In finite sample, $g_{n}(\theta_{0})\neq0$ because $g_{n}(\theta)\neq0$ for all $\theta$ almost surely,  and this causes extra variations through the terms in \eqref{Twostep_Expansion} and \eqref{Onestep_Expansion}. Similar to the Windmeijer correction, by taking into account for these (asymptotically negligible) terms in estimating the variance we can make more accurate inference.

Since $D_{n}=O_{p}(n^{-1/2})$, the terms in \eqref{Onestep_Expansion} multiplied by $D_{n}$ are $O_{p}(n^{-1})$, which is the same order as the remainder term. Thus, considering those terms in \eqref{Onestep_Expansion} does not necessarily provide finite sample corrections. However, including these terms are critical to getting robustness to misspecification, which is shown in Section \ref{Misspecification}.

The expansion for the one-step GMM is \eqref{T1_Exp1}-\eqref{T1_Exp2} and those terms in \eqref{T1_Exp12} are $O_{p}(n^{-1/2})$. Thus, considering the finite sample variation caused by these terms provides a more accurate variance estimator formula. This correction for the one-step GMM is not considered in Windmeijer (2005) and is new.

The doubly corrected variance estimator of $\sqrt{n}(\hat{\theta}_{2}-\theta_{0})$ is
\begin{equation}
\widehat{V}_{dc}(\hat{\theta}_{2}) =\widehat{V}(\hat{\theta}_{2}) 
+\widehat{D}_{n}\widehat{C}(\hat{\theta}_{1},\hat{\theta}_{2}) +\widehat{C}(\hat{\theta}_{1},\hat{\theta}_{2})^{\prime}\widehat{D}_{n}^{\prime}+\widehat{D}_{n}\widehat{V}_{dc}(\hat{\theta}_{1})\widehat{D}_{n}^{\prime},
\label{Vdc2}
\end{equation}
where 
\begin{align}
\label{Vhat2} \widehat{V}(\hat{\theta}_{2}) =& \left(G_{n}^{\prime}[\Omega_{n}(\hat{\theta}_{1})]^{-1}G_{n}\right)^{-1}\Sigma_{n}(\hat{\theta}_{2},\Omega_{n}(\hat{\theta}_{1}))\left(G_{n}^{\prime}[\Omega_{n}(\hat{\theta}_{1})]^{-1}G_{n}\right)^{-1},\\
\label{Vdc1} \widehat{V}_{dc}(\hat{\theta}_{1}) =& \left(G_{n}^{\prime}W_{n}^{-1}G_{n}\right)^{-1}\Sigma_{n}(\hat{\theta}_{1},W_{n})\left(G_{n}^{\prime}W_{n}^{-1}G_{n}\right)^{-1},\\
\label{Chat} \widehat{C}(\hat{\theta}_{1},\hat{\theta}_{2}) =& \left(G_{n}^{\prime}W_{n}^{-1}G_{n}\right)^{-1}\frac{1}{n}\sum_{i=1}^{n}m_{i}(\hat{\theta}_{1},W_{n})m_{i}(\hat{\theta}_{2},\Omega_{n}(\hat{\theta}_{1}))^{\prime}\left(G_{n}^{\prime}[\Omega_{n}(\hat{\theta}_{1})]^{-1}G_{n}\right)^{-1},
\end{align}
and
\begin{align}
\label{Sigman} \Sigma_{n}(\theta,\Xi_{n}(\phi)) =& \frac{1}{n}\sum_{i=1}^{n}m_{i}(\theta,\Xi_{n}(\phi))m_{i}(\theta,\Xi_{n}(\phi))^{\prime},\\
\notag m_{i}(\theta,\Xi_{n}(\phi)) =& G_{n}^{\prime}[\Xi_{n}(\phi)]^{-1}g(X_{i},\theta) + G(X_{i})^{\prime}[\Xi_{n}(\phi)]^{-1}g_{n}(\theta)\\
\notag & -G_{n}^{\prime}[\Xi_{n}(\phi)]^{-1}\Xi(X_{i},\phi)[\Xi_{n}(\phi)]^{-1}g_{n}(\theta),\\
\notag \Xi_{n}(\phi) =& \frac{1}{n}\sum_{i=1}^{n}\Xi(X_{i},\phi).
\end{align}
When $\Xi_{n}(\phi)=\Xi(X_{i},\phi)=I$, the last term of $m(\theta,\Xi_{n}(\phi))$ drops. Note that $G(X_{i})$ and $\Xi(X_{i},\phi)$ in $m_{i}(\theta,\Xi_{n}(\phi))$ are not centered because the FOCs hold evaluated at $(\hat{\theta}_{2},\Omega_{n}(\hat{\theta}_{1}))$ and $(\hat{\theta}_{1},W_{n})$, respectively. 

The doubly corrected variance estimator for the two-step GMM, $\widehat{V}_{dc}(\hat{\theta}_{2})$, provides the same order of finite sample correction as the Windmeijer correction, $\widehat{V}_{w}(\hat{\theta}_{2})$. The standard error is obtained by taking the diagonal elements of $\sqrt{\widehat{V}_{dc}(\hat{\theta}_{2})/n}$.

The doubly corrected variance estimator for the one-step GMM, $\widehat{V}_{dc}(\hat{\theta}_{1})$, accounts for the variations up to the order of $O_{p}(n^{-1/2})$ in the expansion \eqref{T1_Exp1}-\eqref{T1_Exp2}. This correction is not considered in Windmeijer (2005). The standard error is obtained by taking the diagonal elements of $\sqrt{\widehat{V}_{dc}(\hat{\theta}_{1})/n}$.

\section{Robustness to Misspecification\label{Misspecification}}

The variance estimators considered so far, the doubly corrected, the Windmeijer corrected, and the conventional, are consistent for the asymptotic variance of $\sqrt{n}(\hat{\theta}_{2}-\theta_{0})$ under correct specification, $E[g(X_{i},\theta_{0})]=0$. In words, correct specification means that an over-identified model exactly holds at a unique parameter value $\theta_{0}$, but this may be too restrictive in reality. Indeed, the sample moment condition does not hold for any finite sample size $n$ almost surely if the model is over-identified, i.e., $g_{n}(\hat{\theta})\neq0$, provided that at least one of the moments is continuously distributed. Thus, it is reasonable to view the assumed moment condition model as the best-approximating model and to allow for possible misspecification. 

Under (global) misspecification, which is defined as 
\begin{equation}
E[g(X_{i},\theta)]=\delta\mathbf{(\theta)}\neq0,~\forall\theta\in
\Theta,\label{Eqn_global_miss_moment}
\end{equation}
where $\delta(\theta)$ is a vector of constants and $\Theta$ is the parameter space, the GMM estimator is consistent for the pseudo-true value, which is defined as the unique minimizer of the population GMM criterion given the weight matrix (Hall and Inoue, 2003). In addition, the asymptotic variance has more terms that are assumed away under correct specification. Thus, the conventional variance estimators are no longer consistent under misspecification. Lee (2014) proposes variance estimators for the one-step and two-step GMM under misspecification. Hansen and Lee (2019) propose a similar robust variance estimator for the iterated GMM. These variance estimators are shown to be consistent regardless of misspecification and they are referred to as the misspecification-robust variance estimator, hereinafter. 

Since the misspecification-robust variance estimators contain additional terms that are not present in the conventional variance estimator, it has been generally conjectured less accurate than the conventional variance estimator under correct specification. We show that this conjecture is not true by showing that the doubly corrected variance estimator $\widehat{V}_{dc}(\hat{\theta}_{2})$ is the misspecification-robust variance estimator.

The robustness of $\widehat{V}_{dc}(\hat{\theta}_{2})$ holds for the following reasons. Recall that the formulas for $\widehat{V}_{dc}(\hat{\theta}_{2})$
and $\widehat{V}_{w}(\hat{\theta}_{2})$ are given by
\begin{align*}
\widehat{V}_{dc}(\hat{\theta}_{2}) =&\widehat{V}(\hat{\theta}_{2}) 
+\widehat{D}_{n}\widehat{C}(\hat{\theta}_{1},\hat{\theta}_{2}) +\widehat{C}(\hat{\theta}_{1},\hat{\theta}_{2})^{\prime}\widehat{D}_{n}^{\prime}+\widehat{D}_{n}\widehat{V}_{dc}(\hat{\theta}_{1})\widehat{D}_{n}^{\prime},\\
\widehat{V}_{w}(\hat{\theta}_{2}) =&\widetilde{V}(\hat{\theta}_{2}) 
+\widehat{D}_{n}\widetilde{V}(\hat{\theta}_{2}) +\widetilde{V}(\hat{\theta}_{2})\widehat{D}_{n}^{\prime}+\widehat{D}_{n}\widetilde{V}(\hat{\theta}_{1})\widehat{D}_{n}^{\prime}.
\end{align*}
The correction term $\widehat{D}_{n}$ corrects for the bias in the variance due to using the weight matrix $[\Omega_{n}(\hat{\theta}_{1})]^{-1}$ rather than $[\Omega_{n}(\theta_{1})]^{-1}$. Since both $\widehat{V}_{dc}(\hat{\theta}_{2})$ and $\widehat{V}_{w}(\hat{\theta}_{2})$ have $\widehat{D}_{n}$, this bias is corrected in both variance estimators. What is not accounted for in the Windmeijer corrected variance estimator is the additional variations in the sample Jacobian $G_{n}$ and the sample weight matrices $W_{n}$ and $\Omega_{n}(\theta_{1})$. These variations are asymptotically negligible under correct specification but become the first-order under misspecification. Our doubly corrected variance estimator $\widehat{V}_{dc}(\hat{\theta}_{2})$ accounts for these variations.

To formally show that $\widehat{V}_{dc}(\hat{\theta}_{2})$ is consistent for the asymptotic variance under misspecification, we introduce some definitions. Define the one-step and two-step GMM (pseudo-) true values as
\begin{align}
\theta_{1} &= \argmin_{\theta \in \Theta}E[g(X_{i},\theta)]'W^{-1}E[g(X_{i},\theta)],\\
\label{PT-gmm2}\theta_{2}&= \argmin_{\theta \in \Theta}E[g(X_{i},\theta)]'[\Omega(\theta_{1})]^{-1}E[g(X_{i},\theta)].
\end{align}
In general $\theta_{1}\neq \theta_{2}$ but $\theta_{1}=\theta_{2}=\theta_{0}$ under correct specification. Write $g_{j}= E[g(X_{i},\theta_{j})]$ and
$\Omega_{j} = \Omega(\theta_{j})$ for $j=1,2$. (Global) misspecification implies that $g_{n}(\theta_{j})=O_{p}(1)$ for $j=1,2$.

The expansion of the GMM estimators under misspecification is quite similar to those under correct specification, except that we need to allow for different pseudo-true values for the one-step and two-step GMM and the moment condition evaluated at the pseudo-true value is not equal to zero. 

Consider the FOC of the two-step GMM \eqref{twofoc}. By expanding $g_{n}(\hat{\theta}_{2})$ around $\theta_{2}$ and $\Omega_{n}(\hat{\theta}_{1})$ around $\theta_{1}$, we can write
\begin{align}
\notag \sqrt{n}(\hat{\theta}_{2}-\theta_{2})   =&-\left\{  G_{n}^{\prime}%
[\Omega_{n}(\theta_{1})]^{-1}G_{n}\right\}  ^{-1}G_{n}^{\prime}[\Omega_{n}(\theta_{1})]^{-1}\sqrt{n}g_{n}(\theta_{2}) +D_{n}^{*} \sqrt{n}(\hat{\theta}_{1}-\theta_{1})+R^{*}_{n}\\
\label{Eq_2step_Mis}   =&\sqrt{n}(\tilde{\theta}_{2}^{*}-\theta_{2})+D_{n}^{*}  \sqrt{n}(\hat{\theta}_{1}-\theta_{1})+R^{*}_{n},
\end{align}
where $\tilde{\theta}_{2}^{*}$ is defined as
\begin{equation}
\tilde{\theta}_{2}^{*}=\arg\min_{\theta\in\Theta}~g_{n}(\theta)^{\prime}%
[\Omega_{n}(\theta_{1})]^{-1}g_{n}(\theta),
\label{Infeasible_2GMM_Mis}
\end{equation}
and 
\begin{align*}
D_{n}^{*}&=F_{1n}^{*} + F_{2n}^{*},\\
F_{1n}^{*} &  =-\left.  \frac{\partial\left\{  G_{n}^{\prime}%
	[\Omega_{n}(\theta)]^{-1}G_{n}\right\}  ^{-1}}{\partial\theta^{\prime}%
}\right\vert _{\theta=\theta_{1}}G_{n}^{\prime}[\Omega_{n}(\theta
_{1})]^{-1}g_{n}(\theta_{2}),\\
F_{2n}^{*} &  =-\left\{  G_{n}^{\prime}[\Omega_{n}(\theta_{1})]^{-1}G_{n}\right\}  ^{-1}\left.  \frac{\partial G_{n}^{\prime}%
	[\Omega_{n}(\theta)]^{-1}g_{n}(\theta_{2})}{\partial\theta^{\prime}%
}\right\vert _{\theta=\theta_{1}},
\end{align*}
and $R_{n}^{*}$ is the remainder term of order $O_{p}(n^{-1/2}\|g_{n}(\theta_{2})\|)$ (this and the order of other terms are formally justified in the proof of Theorem 1). Since $D_{n}^{*} = O_{p}(\|g_{n}(\theta_{2})\|)$, the order of finite sample correction depends on the degree of misspecification, from being $O_{p}(n^{-1/2})$ under correct specification to $O_{p}(1)$ under (global) misspecification. Note that both $\sqrt{n}(\tilde{\theta}_{2}^{*}-\theta_{2})$ and $D_{n}^{*}  \sqrt{n}(\hat{\theta}_{1}-\theta_{1})$ in \eqref{Eq_2step_Mis} are $O_{p}(1)$ under misspecification and this will alter the first-order asymptotic variance. 

Using the population FOC $G'\Omega_{j}^{-1}g_{j}=0$ for $j=1,2$, the FOC of the infeasible two-step GMM \eqref{Infeasible_2GMM_Mis} can be expanded as
\begin{align}
\notag \sqrt{n}(\tilde{\theta}_{2}^{*}-\theta_{2}) =&-\left\{  G_{n}^{\prime}[\Omega_{n}(\theta_{1})]^{-1}G_{n}\right\}^{-1}\left\{G_{n}^{\prime}[\Omega_{n}(\theta_{1})]^{-1}\sqrt{n}\left(g_{n}(\theta_{2})-g_{2}\right)\right.\\
\label{Eq_Infeasible_Mis} &\left. + \sqrt{n}\left(G_{n}-G\right)^{\prime}[\Omega_{n}(\theta_{1})]^{-1}g_{2} - G^{\prime}\Omega_{1}^{-1}\sqrt{n}\left(\Omega_{n}(\theta_{1})-\Omega_{1}\right)[\Omega_{n}(\theta_{1})]^{-1}g_{2}\right\}.
\end{align}
The FOC of the one-step GMM can be expanded similarly:
\begin{align}
\notag \sqrt{n}(\hat{\theta}_{1}-\theta_{1})=&-\left\{  G_{n}^{\prime}W_{n}^{-1}G_{n}\right\}  ^{-1}\left[G_{n}^{\prime}W_{n}^{-1}\sqrt{n}\left(g_{n}(\theta_{1})-g_{1}\right)\right.\\
&+\left. \sqrt{n}(G_{n}-G)^{\prime}W_{n}^{-1}g_{1} - G^{\prime}W^{-1}\sqrt{n}\left(W_{n}-W\right)W_{n}^{-1}g_{1}\right].
\label{Eq_1step_Mis}
\end{align}
The expansions \eqref{Eq_Infeasible_Mis} and \eqref{Eq_1step_Mis} are misspecification-robust versions of \eqref{Eq_Mean_Value_Infeasible_Two_Step} and \eqref{Eq_Mean_Value_First_Step}, allowing for different probability limits of the one-step and two-step GMM estimators and taking into account for the misspecification (over-identification) bias. Under correct specification, $g_{1}=g_{2}=0$ and \eqref{Eq_Infeasible_Mis} and \eqref{Eq_1step_Mis} coincide with \eqref{Eq_Mean_Value_Infeasible_Two_Step} and \eqref{Eq_Mean_Value_First_Step}.

Now we list assumptions for the main result. 

\begin{assumption}\
	\begin{enumerate}
		\item[(i)] $\theta_{j}$ is unique and is in the interior of the parameter space $\Theta$ for $j=1,2$
		\item[(ii)] $X_{1},\cdots,X_{n}$ are i.i.d. 		
		\item[(iii)] $W$ and $\Omega(\theta_{1})$ are nonsingular
		\item[(iv)] $G$ is full column rank
		\item[(v)] $E[\|g(X_{i},\theta_{1})\|^{4}]<\infty$, $E[\|g(X_{i},\theta_{2})\|^{2}]<\infty$, $E[\|G(X_{i})\|^{4}]<\infty$, $E[\|W(X_{i})\|^{2}]<\infty$
	\end{enumerate}
\end{assumption}

Assumption 1 is mild regularity conditions for consistency and asymptotic normality of the one-step and two-step linear GMM estimators allowing for global misspecification. For misspecified models, Hall and Inoue (2003) provide a list of conditions for one-step and two-step GMM in the time series context. Hansen and Lee (2019) provide a list of conditions for one-step and iterated GMM under the i.n.i.d. and clustered sampling.   

The following theorem shows that the doubly corrected variance estimators of the one-step and two-step linear GMM are consistent for the asymptotic variance matrices under misspecification. The proof is given in the Appendix A.

\begin{theorem}
	Suppose that Assumption 1 holds. As $n\rightarrow\infty$, for $j=1,2$,
	\begin{equation*}
\label{TMRm}
\sqrt{n}(\hat{\theta}_{j}-\theta_{j})\xrightarrow{d}N(0,V_{j})
\end{equation*}
	and 
	\begin{equation*}
\label{TMRv}
\widehat{V}_{dc}(\hat{\theta}_{j})\xrightarrow{p}V_{j}.
\end{equation*}
	\label{TMR}
\end{theorem}

Theorem \ref{TMR} holds regardless of whether the model is correctly specified or not. Thus, $\widehat{V}_{dc}(\hat{\theta}_{j})$ for $j=1,2$, provides a finite sample correction under correct specification and it remains consistent under misspecification. In contrast, the Windmeijer corrected variance estimator, $\widehat{V}_{w}(\hat{\theta}_{2})$, is not first-order consistent under misspecification. Assuming the linearity of moment condition, the proofs of Lemma 2 and Theorem 1 in Appendix A can be used to rigorously justify the stochastic orders of the higher-order terms mentioned in the previous sections. 

For nonlinear models, $\widehat{V}_{dc}(\hat{\theta}_{j})$ formula can be adjusted by considering the second derivative of the moment function, which coincides with the misspecification-robust formula of Lee (2014) and Hansen and Lee (2019). The same consistency result with Theorem 1 are shown in those papers, but stronger assumptions on the moment/Jacobian processes and the compact parameter space are required to use the uniform law of large numbers.

Theorem \ref{TMR} also implies that the nonparametric i.i.d. bootstrap $t$ test and confidence intervals (CIs) based on the GMM $t$ statistic studentized with the doubly corrected standard error automatically achieve higher-order refinements over the asymptotic $t$ test and CIs regardless of misspecification (Lee, 2014). In contrast, those bootstrap $t$ test and CIs based on the GMM $t$ statistic studentized with the conventional or the Windmeijer standard error require an additional recentering procedure in resampling to correct for the over-identification bias to achieve higher-order refinements, see Hall and Horowitz (1996) and Andrews (2002). Furthermore, the conventional nonparametric i.i.d. bootstrap procedure for GMM is not valid under misspecification. Thus, our doubly corrected variance estimator formula provides a very convenient way to get more accurate but also robust bootstrap tests and CIs.

\noindent

\bigskip
\noindent\textbf{Remark 1 (Stochastic expansion under local misspecification)} From the way it is constructed, we can directly see that the doubly corrected variance formulas provides a finite-sample variance correction with the same argument as Windmeijer (2005). It is expected to work the best when the sample moment condition is large. Since a nonzero sample moment condition can also be due to a locally misspecified moment condition, one can seek an additional justification of the double correction by deriving the stochastic expansions of the GMM estimators under such a sequence. Appendix B provides formal stochastic expansions of the one-step and two-step GMM estimators by allowing the population moment condition evaluated at the true value is $E[g(X_{in},\theta_{0})]=\delta/\sqrt{n}$ for some $\delta\neq0$. In Theorems \ref{thm:onestep_stochastic} and \ref{thm:stochastic_expansion} in Appendix B, we discuss how the first-order terms $D_{n}^{*}$, \eqref{Eq_Infeasible_Mis}, and \eqref{Eq_1step_Mis} in the double correction are related to the higher-order terms under local misspecification. Interestingly, our analysis reveals that the double correction effectively estimates the (co)variances of higher-order terms that depend on $\delta$ and thus is fully robust to additional variations due to local misspecification. In contrast, the Windmeijer corrected variance estimator only partially considers the higher-order terms that depend on $\delta$, making it only partly robust to local misspecification. Finally, our stochastic expansions show that both the doubly corrected and the Windmeijer corrected variance estimators are not higher-order variance estimators which would estimate additional terms up to $O(n^{-1})$. For the general treatment of the stochastic expansion, see Rothenberg (1984). Newey and Smith (2004) derive the stochastic expansion of GMM and the generalized empirical likelihood (GEL) estimators under correct specification.

\bigskip

\noindent\textbf{Remark 2 (Weight matrix)} The main results hold if we replace $\Omega_{n}(\theta)$ with the centered weight matrix
\begin{equation}
\label{Weight_Centered}
\Omega^{c}_{n}(\theta) = \frac{1}{n}\sum_{i=1}^{n}\left(g(X_{i},\theta)-g_{n}(\theta)\right)\left(g(X_{i},\theta)-g_{n}(\theta)\right)^{\prime},
\end{equation}
with some specifics need to be modified accordingly. Specifically, the two-step GMM pseduo-true value $\theta_{2}$ defined in \eqref{PT-gmm2} is now defined with $\Omega^{c}_{n}(\theta_{1})$. In addition, the derivative of the centered weight matrix is
\begin{equation*}
\Upsilon_{j}^{c}(\theta) = \frac{1}{n}\sum_{i=1}^{n}(g(X_{i},\theta)-g_{n}(\theta))\left(\frac{\partial g(X_{i},\theta)}{\partial\theta_{[j]}}-\frac{\partial g_{n}(\theta)}{\partial\theta_{[j]}}\right)'
\end{equation*}
so that 
\begin{equation*}
\frac{\partial\Omega_{n}^{c}(\theta)}{\partial\theta_{[j]}}=\Upsilon^{c}
_{j}(\theta)+\Upsilon_{j}^{c\prime}(\theta).
\end{equation*}
The centered weight matrix is consistent for the asymptotic variance matrix of the moment equation under misspecification. Hansen (2020) recommends using the centered weight matrix for this reason. Hall (2000) shows that the GMM over-identification test statistic with a centered heteroskedasticity-and-autocorrelation-consistent (HAC) weight matrix leads to more powerful tests in the time series setting.

\section{Iterated GMM and Continuously Updating GMM}\label{Section_iteragedGMM}

Both the Windmeijer and our double correction correct for the extra variation due to the weight matrix being evaluated at an estimate rather than the true value. A natural question is whether similar finite sample corrections can be obtained for other GMM estimators, namely the iterated GMM of B. Hansen and Lee (2019) and the continuously-updating (CU) GMM of L. Hansen, Heaton, and Yaron (1996). We show that the answer is yes for the iterated GMM and the double correction formula is the same as the misspecification-robust formula. For the CU GMM, the answer is negative.

Assume correct specification. The iterated GMM estimator is obtained by iterating the two-step efficient GMM estimator until convergence. By iteration the dependence of the final estimator on the previous step estimators disappears. The FOC is given by
\begin{equation}
0=G_{n}'[\Omega_{n}(\hat{\theta})]^{-1}g_{n}(\hat{\theta})
\end{equation}
where $\hat{\theta}$ is the iterated GMM. Assume that $g_{n}(\theta_{0})=O_{p}(n^{-1/2})$ and $\hat{\theta} -\theta_{0} = O_{p}(n^{-1/2})$ whose sufficient conditions are provided in Hansen and Lee (2019). By applying the first-order Taylor expansion around $\theta_{0}$ to $g_{n}(\hat{\theta})$ and $\Omega_{n}(\hat{\theta})$ sequentially
\begin{align}
\notag \sqrt{n}(\hat{\theta}-\theta_{0})=& -\left\{G_{n}'[\Omega_{n}(\theta_{0})]^{-1}G_{n}\right\}^{-1}G_{n}^{\prime}[\Omega_{n}(\theta_{0})]^{-1}\sqrt{n}g_{n}(\theta_{0})+ D_{n}\sqrt{n}(\hat{\theta}-\theta_{0}) + O_{p}(n^{-1})
\end{align}
and thus
\begin{align}
\label{Eq_iterated0}
\sqrt{n}(\hat{\theta}-\theta_{0}) =& -\left\{G_{n}'[\Omega_{n}(\theta_{0})]^{-1}G_{n}(I_{k} - D_{n})\right\}^{-1}G_{n}^{\prime}[\Omega_{n}(\theta_{0})]^{-1}\sqrt{n}g_{n}(\theta_{0}) + O_{p}(n^{-1}).
\end{align}
Windmeijer (2000) proposes a finite sample corrected variance estimator based on the expansion \eqref{Eq_iterated0}. We proceed one additional step. By further expanding to take into account for the over-identification bias, we have
\begin{align}
\label{Eq_iterated1}\sqrt{n}(\hat{\theta}-\theta_{0}) &= -\left\{G_{n}'[\Omega_{n}(\theta_{0})]^{-1}G_{n}(I_{k} - D_{n})\right\}^{-1}\left[G^{\prime}\Omega^{-1}\sqrt{n}g_{n}(\theta_{0})\right.\\
\label{Eq_iterated2} & \left. + \sqrt{n}(G_{n}-G)^{\prime}\Omega^{-1}g_{n}(\theta_{0}) - G^{\prime}\Omega^{-1}\sqrt{n}\left(\Omega_{n}(\theta_{0})-\Omega\right)\Omega^{-1}g_{n}(\theta_{0})\right]+  O_{p}(n^{-1}).
\end{align}
Since the remainder term is $O_{p}(n^{-1})$, by estimating the variance of the terms in \eqref{Eq_iterated1}-\eqref{Eq_iterated2} up to $O_{p}(n^{-1/2})$ we can get the same order of finite sample correction with the doubly corrected two-step GMM variance estimator.

The doubly corrected variance estimator for the iterated GMM is 
\begin{align}
\widehat{V}_{dc}(\hat{\theta}) &= \{G_{n}^{\prime}[\Omega_{n}(\hat{\theta})]^{-1}G_{n}(I_{k}-\widehat{D}_{n})\}^{-1}\Sigma_{n}(\hat{\theta},\Omega_{n}(\hat{\theta}))\{G_{n}^{\prime}[\Omega_{n}(\hat{\theta})]^{-1}G_{n}(I_{k}-\widehat{D}_{n})\}^{-1\prime},
\end{align}
where $\Sigma_{n}(\hat{\theta},\Omega_{n}(\hat{\theta}))$ is defined in \eqref{Sigman} and $\widehat{D}_{n}$ is evaluated at $\hat{\theta}$. Not surprisingly, this formula is identical to the misspecification-robust variance estimator for the iterated GMM of Hansen and Lee (2019). The finite sample corrected formula suggested by Windmeijer (2000) is
\begin{equation}
\widehat{V}_{w}(\hat{\theta}) = (I_{k}-\widehat{D}_{n})^{-1}\left(G_{n}'[\Omega_{n}(\hat{\theta})]^{-1}G_{n}\right)^{-1}(I_{k}-\widehat{D}_{n})^{-1'}.
\end{equation}

On the other hand, a similar finite sample correction may not be obtained for the CU GMM. Windmeijer (2005) showed that if the derivative of the moment function is a function of the parameter, then the proposed formula would not necessarily give finite sample corrections. The same argument applied to CU GMM. Let $\hat{\theta}$ be the CU GMM estimator. For simplicity, let $k=1$ so that $\theta$ is scalar. The FOC is 
\begin{equation}
\label{FOC_CUE}
0 = \left(G_{n}-\frac{1}{2}g_{n}(\hat{\theta})'[\Omega_{n}(\hat{\theta})]^{-1}\left.\frac{\partial\Omega_{n}(\theta)}{\partial\theta}\right|_{\theta=\hat{\theta}}\right)'[\Omega_{n}(\hat{\theta})]^{-1}g_{n}(\hat{\theta}).
\end{equation}
This shows that even when the moment function is linear, the effective Jacobian term in the FOC still depends on the parameter. Thus, the misspecification-robust variance formula for CU GMM does not necessarily provide a finite sample correction under correct specification. Since GEL estimators have similar non-linear FOC even with linear moment functions, we expect similar conclusions. 

\section{Examples}\label{Section_Examples}

\subsection{Cross-sectional IV}\label{Section_Examples_IV}

Consider the linear IV model $y_{i} = X_{i}^{\prime} \theta + e_{i}$ with the moment conditions $E[Z_{i} e_{i}]=0$. The two-stage least squares (2SLS) estimator is given by 
\begin{equation}
\hat{\theta}_{1}=(X^{\prime} Z (Z^{\prime}Z)^{-1} Z^{\prime} X)^{-1} X^{\prime} Z (Z^{\prime}Z)^{-1} Z^{\prime} Y
\end{equation}
where $Y = [y_{1}, \cdots, y_{n}]^{\prime}, X = [X_{1}, \cdots, X_{n}]^{\prime}$, and $Z = [Z_{1}, \cdots, Z_{n}]^{\prime}$ are $n\times 1$, $n\times k$, and $n\times q$ data matrices. Using the 2SLS as the preliminary estimator, the two-step efficient GMM estimator is given by
\begin{equation}
\hat{\theta}_{2}=(X^{\prime} Z \widehat{\Omega}_{1}^{-1} Z^{\prime} X)^{-1} X^{\prime} Z \widehat{\Omega}_{1}^{-1} Z^{\prime} Y
\end{equation}
where 
\begin{align*}
\widehat{\Omega}_{1}&=\frac{1}{n}\sum_{i=1}^{n} Z_{i} Z_{i}^{\prime} \hat{e}_{1i}^{2},\\
\hat{e}_{1i} &= y_{i} - X_{i}^{\prime} \hat{\theta}_{1}.
\end{align*}
Also define $\hat{e}_{2i} = y_{i} - X_{i}^{\prime} \hat{\theta}_{2}$ and the $n\times 1$ residual vector $\hat{e}_{j} = Y-X\hat{\theta}_{j}$ for $j=1,2$. 

The doubly corrected variance estimators of the 2SLS and two-step GMM are 
\begin{align}
\widehat{V}_{dc}(\hat{\theta}_{1}) =& \left(\frac{1}{n} X^{\prime}Z (Z^{\prime}Z)^{-1} Z^{\prime} X\right)^{-1}\frac{1}{n}\sum_{i=1}^{n}\hat{m}_{1i}\hat{m}_{1i}^{\prime}\left(\frac{1}{n} X^{\prime}Z (Z^{\prime}Z)^{-1} Z^{\prime} X\right)^{-1},\\
\widehat{V}_{dc}(\hat{\theta}_{2}) =&\widehat{V}(\hat{\theta}_{2}) 
+\widehat{D}_{n}\widehat{C}(\hat{\theta}_{1},\hat{\theta}_{2}) +\widehat{C}(\hat{\theta}_{1},\hat{\theta}_{2})^{\prime}\widehat{D}_{n}^{\prime}+\widehat{D}_{n}\widehat{V}_{dc}(\hat{\theta}_{1})\widehat{D}_{n}^{\prime}, 
\end{align}
where 
\begin{align*}
\widehat{V}(\hat{\theta}_{2}) =& \left(\frac{1}{n^2} X^{\prime}Z \widehat{\Omega}_{1}^{-1} Z^{\prime} X\right)^{-1}\left(\frac{1}{n}\sum_{i=1}^{n}\hat{m}_{2i}\hat{m}_{2i}^{\prime}\right)\left(\frac{1}{n^2} X^{\prime}Z \widehat{\Omega}_{1}^{-1} Z^{\prime} X\right)^{-1},\\
\widehat{C}(\hat{\theta}_{1},\hat{\theta}_{2}) =& \left(\frac{1}{n} X^{\prime}Z (Z^{\prime}Z)^{-1} Z^{\prime} X\right)^{-1}\left(\frac{1}{n}\sum_{i=1}^{n}\hat{m}_{1i}\hat{m}_{2i}^{\prime}\right)\left(\frac{1}{n^2} X^{\prime}Z \widehat{\Omega}_{1}^{-1} Z^{\prime} X\right)^{-1},\\
\widehat{D}_{n} =&\frac{2}{n}\left( X^{\prime}Z \widehat{\Omega}_{1}^{-1} Z^{\prime} X\right)^{-1}  X^{\prime}Z \widehat{\Omega}_{1}^{-1} \sum_{i=1}^{n} Z_{i} \left(\hat{e}_{1i} Z_{i}^{\prime}\widehat{\Omega}_{1}^{-1}  Z^{\prime} \hat{e}_{2}\right)  X_{i}^{\prime} ,\\
\hat{m}_{1i} =& X^{\prime} Z (Z^{\prime}Z)^{-1}Z_{i} \hat{e}_{1i} + X_{i} Z_{i}^{\prime} (Z^{\prime}Z)^{-1}  Z^{\prime} \hat{e}_{1} - X^{\prime} Z (Z^{\prime}Z)^{-1}Z_{i} Z_{i}^{\prime} (Z^{\prime}Z)^{-1}  Z^{\prime} \hat{e}_{1},\\
\hat{m}_{2i} =& \frac{1}{n} X^{\prime}Z \widehat{\Omega}_{1}^{-1} Z_{i} \hat{e}_{2i} +   \frac{1}{n}X_{i} Z_{i}^{\prime} \widehat{\Omega}_{1}^{-1}  Z^{\prime} \hat{e}_{2}- \frac{1}{n^{2}} X^{\prime}Z \widehat{\Omega}_{1}^{-1} Z_{i} Z_{i}^{\prime}  \hat{e}_{1i}^{2} \widehat{\Omega}_{1}^{-1}  Z^{\prime} \hat{e}_{2}.
\end{align*}

It is worth observing that the doubly corrected variance estimator $\widehat{V}_{dc}(\hat{\theta}_{2})$ reduces to the Windmeijer corrected one $\widehat{V}_{w}(\hat{\theta}_{2})$ if (i) the last two terms in $\hat{m}_{2i}$ and $\hat{m}_{1i}$ are ignored and (ii) $\hat{e}_{1i}$ replaces $\hat{e}_{2i}$ in $\hat{m}_{2i}$. By (i) and (ii), the variance estimators $\widehat{V}(\hat{\theta}_{2})$ and $\widehat{V}_{dc}(\hat{\theta}_{1})$ reduce to conventional ones $\widetilde{V}(\hat{\theta}_{2})$ and $\widetilde{V}(\hat{\theta}_{1})$, and $\widehat{C}(\hat{\theta}_{1},\hat{\theta}_{2})$ becomes $\widetilde{V}(\hat{\theta}_{2})$. In general, however, $ \widehat{V}_{dc}(\hat{\theta}_{2})\neq \widehat{V}_{w}(\hat{\theta}_{2})$ because $Z^{\prime}\hat{e}_{j} \neq0$ for $j=1,2$, so the last two terms of $\hat{m}_{ji}$ are non-zero. Furthermore, it is critical (and reasonable) to use $\hat{e}_{2i}$ in $\hat{m}_{2i}$ to get robustness under misspecification. 

The iterated GMM estimator is obtained as follows. Let $\hat{\theta}_{0}$ be any initial value. The $s$-step GMM estimator for $s\geq1$ is given by
\begin{equation}
\hat{\theta}_{s}=(X^{\prime} Z \widehat{\Omega}_{s-1}^{-1} Z^{\prime} X)^{-1} X^{\prime} Z \widehat{\Omega}_{s-1}^{-1} Z^{\prime} Y,
\end{equation}
where 
\begin{equation*}
\widehat{\Omega}_{s-1} = \frac{1}{n}\sum_{i=1}^{n} Z_{i} Z_{i}^{\prime} (y_{i}-X_{i}'\hat{\theta}_{s-1})^{2}.
\end{equation*}
We iterate the $s$-step GMM estimator until convergence given a preset tolerance $\epsilon$, i.e. $\|\hat{\theta}_{s}-\hat{\theta}_{s-1}\|<\epsilon$ to obtain the iterated GMM estimator $\hat{\theta}$. The residuals are $\hat{e}_{i} = y_{i} - X_{i}^{\prime} \hat{\theta}$. Also let $\hat{e} = Y-X\hat{\theta}$ be the $n\times1$ residual vector.

The doubly corrected variance estimator is
\begin{align}
\widehat{V}_{dc}(\hat{\theta}) &= \widehat{H}^{-1}\left( \frac{1}{n}\sum_{i=1}^{n}\hat{m}_{i}\hat{m}_{i}^{\prime}\right) \widehat{H}^{-1\prime},\\
\notag \widehat{H} &=\frac{1}{n^2} X^{\prime}Z\widehat{\Omega}^{-1} Z^{\prime} X - \frac{2}{n^{3}} X^{\prime}Z \widehat{\Omega}^{-1} \sum_{i=1}^{n} Z_{i} \left(\hat{e}_{i} Z_{i}^{\prime}  \widehat{\Omega}^{-1}  Z^{\prime} \hat{e}\right) X_{i}^{\prime} ,\\
\notag \hat{m}_{i} &=\frac{1}{n} X^{\prime}Z \widehat{\Omega}^{-1} Z_{i} \hat{e}_{i} +  \frac{1}{n}X_{i} Z_{i}^{\prime}  \widehat{\Omega}^{-1}  Z^{\prime} \hat{e} - \frac{1}{n^{2}} X^{\prime}Z \widehat{\Omega}^{-1}Z_{i} Z_{i}^{\prime}  \hat{e}_{i}^{2} \widehat{\Omega}^{-1}  Z^{\prime} \hat{e}.
\end{align}
In comparison, the Windmeijer corrected and the conventional variance estimators are
\begin{align}
\widehat{V}_{w}(\hat{\theta}) &= \widehat{H}^{-1}\left(\frac{1}{n^2} X^{\prime}Z\widehat{\Omega}^{-1} Z^{\prime} X\right) \widehat{H}^{-1'},\\
\widetilde{V}(\hat{\theta})&=\left(\frac{1}{n^2} X^{\prime}Z\widehat{\Omega}^{-1} Z^{\prime} X\right)^{-1}.
\end{align}

\subsection{A Panel Data Model}\label{Section_Examples_Panel}

Consider a panel data model with a scalar regressor
\begin{equation}
y_{it}  =x_{it} \beta + \eta_{i}+v_{it},
\label{panel_model}
\end{equation}
for $i = 1, ..., N$ and $t = 1, ... ,T$ where  $\eta_{i}$ is the unobserved individual effects,  the unknown parameter of interest is $\beta$, and the single regressor $x_{it}$ is predetermined with respect to $v_{it}$ (possibly including lags of the dependent variable), i.e., $E(x_{it}v_{is})=0$ for all $s\geq t$. After first-differencing, 
\begin{equation*}
\Delta y_{it}  =\Delta x_{it}\beta + \Delta v_{it}, \quad t = 2, ... ,T, 
\end{equation*}
the standard approach to estimate $\beta$ is the first differenced GMM (Arellano and Bond (1991) estimator) with the moment conditions $E(Z_{i}^{\prime} \Delta v_{i}) = 0$ where $Z_{i}$ is the $(T-1)$ by $T(T-1)/2$ instrument matrix 
\begin{equation*}
Z_{i}=diag(z_{i2}^{\prime},\cdots,z_{iT}^{\prime})
\end{equation*} with all possible
lagged instruments $z_{it}=(x_{i1},\cdots,x_{it-1})^{\prime}$ for $2\leq t\leq T$ and $\Delta v_{i} = (\Delta v_{i2},\cdots, \Delta v_{iT})^{\prime}$. The total number of observations is $n=N(T-1)$.

Our doubly corrected variance estimator can be used for the model \eqref{panel_model} with additional strictly exogenous, predetermined, or endogenous variables as well as the system GMM estimator (Arellano and Bover (1995) and Blundell and Bond (1998)) by stacking and modifying additional moment conditions into the instrument sets $Z_{i}$. If the panel is unbalanced the instrument matrix can be constructed as described in Arellano and Bond (1991). 

Using the initial weight matrix $\widehat{W}  = n^{-1}\sum_{i=1}^{N}Z_{i}^{\prime} H Z_{i}$, where $H$ is a matrix with $2$'s on the main diagonal, $-1$'s on the first off-diagonals and zero elsewhere, the one-step GMM estimator is given by 
 \begin{equation*}
\hat{\beta}_{1}=(\Delta X^{\prime} Z \widehat{W}^{-1} Z^{\prime} \Delta X)^{-1} \Delta X^{\prime} Z \widehat{W}^{-1}  Z^{\prime} \Delta Y
\end{equation*}
where $Z = (Z_{1}^{\prime}, ..., Z_{N}^{\prime})^{\prime}$ is the instrument matrix, $\Delta Y = (\Delta y_{1}^{\prime}, ..., \Delta y_{N}^{\prime})^{\prime}$, $\Delta X = (\Delta x_{1}^{\prime}, ..., \Delta x_{N}^{\prime})^{\prime}$, $\Delta y_{i} = (\Delta y_{i2}, ..., \Delta y_{iT})'$, and  $\Delta x_{i} = (\Delta x_{i2}, ..., \Delta x_{iT})'$. Note that scaling the weight matrix does not affect the estimator. The doubly corrected variance estimator of $\hat{\beta}_{1}$ is given by 
\begin{align*}
\widehat{V}_{dc}(\hat{\beta}_{1}) &=  n^{2}\left(\Delta X^{\prime}Z \widehat{W}^{-1} Z^{\prime} \Delta X\right)^{-1} \left(\frac{1}{n}\sum_{i=1}^{N}\hat{m}_{1i}\hat{m}_{1i}^{\prime}\right)\left(\Delta X^{\prime}Z \widehat{W}^{-1} Z^{\prime} \Delta X\right)^{-1},\\
\hat{m}_{1i} &=\Delta X^{\prime} Z \widehat{W}^{-1} Z_{i}^{\prime} \Delta \hat{v}_{1i} + \Delta x_{i}^{\prime} Z_{i}  \widehat{W}^{-1} Z^{\prime} \Delta \hat{v}_{1}- \frac{1}{n} \Delta X^{\prime} Z \widehat{W}^{-1}Z_{i}^{\prime} H Z_{i} \widehat{W}^{-1} Z^{\prime} \Delta \hat{v}_{1},
\end{align*}
where $\Delta \hat{v}_{1i} = \Delta y_{i} - \Delta x_{i} \hat{\beta}_{1}$ and $\Delta \hat{v}_{1} = (\Delta \hat{v}_{11}^{\prime}, ..., \Delta \hat{v}_{1N}^{\prime})^{\prime}$. The doubly corrected standard error is obtained by taking the diagonal elements of $\sqrt{\widehat{V}_{dc}(\hat{\beta}_{1})/n}$. In comparison, the conventional variance estimator is given by
\begin{equation*}
\widetilde{V}(\hat{\beta}_{1})=n^{2}\left(\Delta X^{\prime}Z \widehat{W}^{-1} Z^{\prime} \Delta X\right)^{-1} \Delta X^{\prime} Z \widehat{W}^{-1} \widehat{\Omega}_{1} \widehat{W}^{-1}Z'\Delta X \left(\Delta X^{\prime}Z \widehat{W}^{-1} Z^{\prime} \Delta X\right)^{-1}
\end{equation*}
where 
\begin{equation}
\widehat{\Omega}_{1} = \frac{1}{n}\sum_{i=1}^{N} Z_{i}^{\prime} \Delta \hat{v}_{1i} \Delta \hat{v}_{1i}^{\prime} Z_{i}.
\end{equation}

Next, consider the two-step efficient GMM estimator 
\begin{equation*}
\hat{\beta}_{2}=(\Delta X^{\prime} Z \widehat{\Omega}_{1} ^{-1} Z^{\prime} \Delta X)^{-1} \Delta X^{\prime} Z \widehat{\Omega}_{1} ^{-1} Z^{\prime} \Delta Y.
\end{equation*}
Let $\Delta \hat{v}_{2i} = \Delta y_{i} - \Delta x_{i} \hat{\beta}_{2}$ and $\Delta \hat{v}_{2} = (\Delta \hat{v}_{21}^{\prime}, ..., \Delta \hat{v}_{2N}^{\prime})^{\prime}$. The doubly corrected variance estimator of $\hat{\beta}_{2}$ is given by 
\begin{equation*}
\widehat{V}_{dc}(\hat{\beta}_{2}) =\widehat{V}(\hat{\beta}_{2}) 
+\widehat{D}_{n}\widehat{C}(\hat{\beta}_{1},\hat{\beta}_{2}) +\widehat{C}(\hat{\beta}_{1},\hat{\beta}_{2})^{\prime}\widehat{D}_{n}^{\prime}+\widehat{D}_{n}\widehat{V}_{dc}(\hat{\beta}_{1})\widehat{D}_{n}^{\prime},
\end{equation*}
where 
\begin{align*}
\widehat{V}(\hat{\beta}_{2}) &=   n^{2}\left(\Delta X^{\prime}Z \widehat{\Omega}_{1}^{-1} Z^{\prime} \Delta X\right)^{-1} \left(\frac{1}{n}\sum_{i=1}^{N}\hat{m}_{2i}\hat{m}_{2i}^{\prime}\right)\left(\Delta X^{\prime}Z \widehat{\Omega}_{1}^{-1} Z^{\prime} \Delta X\right)^{-1},\\
\widehat{C}(\hat{\beta}_{1},\hat{\beta}_{2}) &=n^{2}\left(\Delta X^{\prime}Z \widehat{W}^{-1} Z^{\prime} \Delta X\right)^{-1} \left(\frac{1}{n}\sum_{i=1}^{N}\hat{m}_{1i}\hat{m}_{2i}^{\prime}\right)\left(\Delta X^{\prime}Z \widehat{\Omega}_{1}^{-1} Z^{\prime} \Delta X\right)^{-1},\\
\hat{m}_{2i} &= \Delta X^{\prime} Z \widehat{\Omega}_{1}^{-1} Z_{i}^{\prime} \Delta \hat{v}_{2i} + \Delta x_{i}^{\prime} Z_{i} \widehat{\Omega}_{1}^{-1}  Z^{\prime} \Delta \hat{v}_{2} - \frac{1}{n} \Delta X^{\prime} Z \widehat{\Omega}_{1}^{-1}  Z_{i}^{\prime} \Delta \hat{v}_{1i} \Delta \hat{v}_{1i}^{\prime} Z_{i}   \widehat{\Omega}_{1}^{-1}  Z^{\prime} \Delta \hat{v}_{2},\\
\widehat{D}_{n} &=\left(\Delta X^{\prime}Z \widehat{\Omega}_{1}^{-1} Z^{\prime} \Delta X\right)^{-1}  \Delta X^{\prime} Z \widehat{\Omega}_{1}^{-1}\\
&\times\frac{1}{n}\sum_{i=1}^{N}  \left(Z_{i}^{\prime}\Delta x_{i}\left(\Delta \hat{v}_{2}'Z\widehat{\Omega}_{1}^{-1}Z_{i}'\Delta \hat{v}_{1i}\right)  + \left(Z_{i}'\Delta \hat{v}_{1i}\right)\left(\Delta \hat{v}_{2}'Z\widehat{\Omega}_{1}^{-1}Z_{i}'\Delta x_{i}\right)\right).
\end{align*}
The doubly corrected standard error is obtained by taking the diagonal elements of $\sqrt{\widehat{V}_{dc}(\hat{\beta}_{2})/n}$. Note that the Windmeijer corrected variance estimator is
\begin{equation*}
\widehat{V}_{w}(\hat{\beta}_{2}) =\widetilde{V}(\hat{\beta}_{2}) 
+\widehat{D}_{n}\widetilde{V}(\hat{\beta}_{2}) +\widetilde{V}(\hat{\beta}_{2})\widehat{D}_{n}^{\prime}+\widehat{D}_{n}\widetilde{V}(\hat{\beta}_{1})\widehat{D}_{n}^{\prime},
\end{equation*}
where
\begin{equation}
\widetilde{V}(\hat{\beta}_{2}) =n^{2}\left(\Delta X^{\prime}Z \widehat{\Omega}_{1}^{-1} Z^{\prime} \Delta X\right)^{-1}.
\end{equation}

Finally, the iterated GMM estimator is given as follows. Let $\hat{\beta}_{0}$ be any initial value. The $s$-step GMM estimator for $s\geq1$ is given by
\begin{equation}
\hat{\beta}_{s}=(\Delta X^{\prime} Z \widehat{\Omega}_{s-1}^{-1} Z^{\prime} \Delta X)^{-1} \Delta X^{\prime} Z \widehat{\Omega}_{s-1}^{-1} Z^{\prime} \Delta Y,
\end{equation}
where 
\begin{equation*}
\widehat{\Omega}_{s-1} = \frac{1}{n}\sum_{i=1}^{N} Z_{i}'(\Delta y_{i} - \Delta x_{i} \hat{\beta}_{s-1})(\Delta y_{i} - \Delta x_{i} \hat{\beta}_{s-1})' Z_{i}.
\end{equation*}
We iterate the $s$-step GMM estimator until convergence given a preset tolerance $\epsilon$, i.e. $\|\hat{\beta}_{s}-\hat{\beta}_{s-1}\|<\epsilon$ to obtain the iterated GMM estimator $\hat{\beta}$. The residuals are $\Delta \hat{v}_{i} = \Delta y_{i} - \Delta x_{i}\hat{\beta}$. Also let $\Delta \hat{v} = (\Delta \hat{v}_{1}^{\prime}, ..., \Delta \hat{v}_{N}^{\prime})^{\prime}$ be the $n\times1$ residual vector. 

The doubly corrected variance estimator for the iterated GMM is given by 
\begin{align*}
\widehat{V}_{dc}(\hat{\beta}) =&  \widehat{H}^{-1} \left(\frac{1}{n}\sum_{i=1}^{N}\hat{m}_{i}\hat{m}_{i}^{\prime}\right) \widehat{H}^{-1'},\\
\widehat{H} =& \frac{1}{n^2} \Delta X^{\prime}Z \widehat{\Omega}^{-1} Z^{\prime} \Delta X\\
& - \frac{1}{n^3} \Delta X^{\prime}Z \widehat{\Omega}^{-1} \left(\sum_{i=1}^{N}\left(Z_{i}'\Delta \hat{v}_{i}\right)\left(\Delta \hat{v}'Z\widehat{\Omega}^{-1}Z_{i}'\Delta x_{i}\right) + Z_{i}'\Delta x_{i}\left(\Delta\hat{v}'Z\widehat{\Omega}^{-1}Z_{i}'\Delta \hat{v}_{i}\right)\right),\\
\hat{m}_{i} =& \frac{1}{n} \Delta X^{\prime}Z \widehat{\Omega}^{-1}Z_{i}^{\prime} \Delta \hat{v}_{i} + \frac{1}{n}\Delta X_{i}^{\prime} Z_{i}  \widehat{\Omega}^{-1}  Z^{\prime} \Delta \hat{v} - \frac{1}{n^2} \Delta X^{\prime}Z \widehat{\Omega}^{-1}Z_{i}^{\prime} \Delta \hat{v}_{i} \Delta \hat{v}_{i}^{\prime} Z_{i} \widehat{\Omega}^{-1}  Z^{\prime} \Delta \hat{v}
\end{align*}
and the doubly corrected standard error is obtained by taking the diagonal elements of $\sqrt{\widehat{V}_{dc}(\hat{\beta})/n}$.

In comparison, the Windmeijer corrected and the conventional variance estimators are
\begin{align}
\widehat{V}_{w}(\hat{\beta}) &= \widehat{H}^{-1}\left(\frac{1}{n^2} \Delta X^{\prime}Z\widehat{\Omega}^{-1} Z^{\prime} \Delta X\right) \widehat{H}^{-1'},\\
\widetilde{V}(\hat{\beta})&=\left(\frac{1}{n^2} \Delta X^{\prime}Z\widehat{\Omega}^{-1} Z^{\prime} \Delta X\right)^{-1}.
\end{align}

\section{Simulation\label{Section: Simulation}}

We investigate the finite sample performance of the doubly corrected standard errors proposed in this paper and provide a thorough comparison with the conventional and the Windmeijer corrected ones under correct specification and misspecification. We consider three different setups: (i) a cross-sectional linear IV model with potentially invalid instruments; (ii) a linear dynamic panel model with a random coefficient; (iii) a linear dynamic panel model with possibly misspecified lag specifications. The number of Monte Carlo simulation is 100,000. 

In an unreported simulation, we also investigate the performance of the estimators with the centered weight matrix \eqref{Weight_Centered}. Since the results are similar and there is no obvious pattern of better performance of the point and variance estimators based on the centered weight matrix compared with those based on the uncentered one (reported) they are not reported.

\subsection{Cross-sectional IV}
\label{subsection:simulation-1}
We use the following simulation design which is a simple linear instrumental variable regression
with a single endogenous regressor. The model to be estimated is
\begin{align}
y_{i}  &  =x_{i}\beta_{0}+e_{i}\nonumber\\
E\left(  z_{i}e_{i}\right)   &  =0 \label{simulation1_model}%
\end{align}
where $x_{i}$ and $\beta_{0}$ are scalar and $z_{i}=(z_{1i},z_{2i}%
,z_{3i},z_{4i})^{\prime}$ is a vector of instrumental variables. We estimate
$\beta_{0}$ by 2SLS (one-step), two-step, and iterated GMM, and calculate the conventional, the Windmeijer corrected, and the doubly corrected standard errors. Our data-generating process (DGP) is%
\begin{align}
y_{i}  &  =x_{i}\beta_{0}+ e_{i},\label{simulation1_dgp}\\ 
x_{i}  &  =\pi_{0}\left(  z_{1i}+z_{2i}+z_{3i}+z_{4i}\right)  +u_{i}%
,\nonumber\\
e_{i} & = \frac{\alpha_{0}}{\sqrt{n}}\left(  z_{1i}-z_{2i}+z_{3i}%
-z_{4i}\right)  +  0.5 u_{i} + \sqrt{1-0.5^2} v_{i}, \nonumber\\
z_{i}  &  \sim N\left(  0,I_{4}\right)  ,\nonumber 
\left(
\begin{array}
[c]{c}%
u_{i}\\
v_{i}%
\end{array}
\right)     \sim N\left(  \left(
\begin{array}
[c]{c}%
0\\
0
\end{array}
\right)  ,\left[
\begin{array}
[c]{cc}%
1 & 0\\
0& z_{1i}^2
\end{array}
\right]  \right)  .\nonumber
\end{align}

We set $\beta_{0}=1$, vary $\alpha_{0}$ from 0 to 1 in steps of 0.2, and set
the first-stage coefficient $\pi_{0}$ so that the first-stage $R^{2}=0.2$. We
set the number of observations as $n=50, 100, 500$.

The parameter $\alpha_{0}$ is the extent that the exclusion condition is locally violated. At $\alpha_{0}=0$, the model is correctly specified. For $\alpha_{0}\neq0$, we find $E (z_{i}e_{i}) = (\alpha_{0}, -\alpha_{0}, \alpha_{0}, -\alpha_{0})^{\prime}/\sqrt{n}  \neq 0$, so the moment condition (\ref{simulation1_model}) fails to hold in finite samples, but it holds asymptotically. 

Means and standard deviations of one-step (2SLS), two-step, and iterated GMM estimators are computed in Table \ref{simulation1_Table}. For all GMM estimators, we report means of the conventional standard errors (se $\hat{\beta}$), the Windmeijer corrected standard errors (se$_{w}$ $\hat{\beta}$), and the doubly corrected standard errors (se$_{dc}$ $\hat{\beta}$).

Table \ref{simulation1_Table} shows that our doubly corrected standard errors remain accurate regardless of misspecification, including the correct specification case ($\alpha_{0} = 0$); the means of corrected standard errors are very close to the standard deviations for all values of $\alpha_{0}$, especially for the two-step and the iterated GMM. Simulation evidence reassures our theory that the doubly corrected standard errors not only take into account variation in the estimation of the weight matrix but also extra variation due to the non-zero sample moments in the over-identified model even under correct specification. Furthermore, our doubly corrected standard errors are the only valid one under misspecification. 

The conventional standard error for the one-step GMM (2SLS) estimator is downward biased under correct specification ($\alpha_{0}=0$) and this bias increases with $\alpha_{0}$. As is well known, the conventional standard error for the two-step is severely downward biased when $\alpha_{0}=0$, and this bias also increases with $\alpha_{0}$. The Windmeijer corrected standard error works well under correct specification, but does not fully account for additional variations when $\alpha_{0}$ is non-zero. The result is similar for the iterated GMM.

It is worth noting that the two-step and iterated GMM point estimates are sensitive to the local violation parameter $\alpha_0$. Interestingly, the point estimate becomes similar to the true value as the degree of misspecification $\alpha_{0}$ increases. This is because, in our DGP, the local misspecification bias (which depends on $\alpha_{0}$) and the higher-order asymptotic bias (which does not depend on $\alpha_{0}$) have opposite signs so that it happens to cancel out each other as $\alpha_{0}$ increases. In contrast, the one-step GMM point estimate varies little across $\alpha_{0}$ because the local misspecification bias is zero.\footnote{The local misspecification bias and the higher-order asymptotic bias can be calculated using the formula in Theorems 2 and 3 in Appendix B.} This is specific to this DGP and cannot be generalized. The size of bias in the point estimate decreases as the sample size gets larger.

\begin{table}[htb]
	\centering
	\vspace{-0.5cm}
	\small
	\begin{tabular}{clcccccc}
	\toprule
	&$\alpha_{0}$ &0 &0.2 &0.4&0.6&0.8&1\\
	\midrule
	$n=50$&$\hat{\beta}_{1}$&1.0833&1.0859&1.0823&1.0833&1.0840&1.0811\\
	&sd $\hat{\beta}_{1}$&0.3229&0.3226&0.3295&0.3417&0.3503&0.3746\\
	&se $\hat{\beta}_{1}$&0.2962&0.2963&0.2985&0.3022&0.3047&0.3113\\
	&se$_{dc}$ $\hat{\beta}_{1}$&0.3346&0.3363&0.3417&0.3516&0.3619&0.3794\\
	\\
	&$\hat{\beta}_{2}$&1.0736&1.0656&1.0517&1.0421&1.0344&1.0243\\
	&sd $\hat{\beta}_{2}$&0.3029&0.3035&0.3142&0.3293&0.3470&0.3753\\
	&se $\hat{\beta}_{2}$&0.2544&0.2549&0.2575&0.2616&0.2646&0.2720\\
	&se$_{w}$ $\hat{\beta}_{2}$&0.2889&0.2900&0.2945&0.3039&0.3133&0.3281\\
	&se$_{dc}$ $\hat{\beta}_{2}$&0.3101&0.3144&0.3231&0.3398&0.3570&0.3813\\
	\\
	&$\hat{\beta}_{iter}$&1.0778&1.0672&1.0519&1.0390&1.0290&1.0154\\
	&sd $\hat{\beta}_{iter}$&0.3026&0.3028&0.3147&0.3312&0.3517&0.3827\\
	&se $\hat{\beta}_{iter}$&0.2513&0.2531&0.2573&0.2631&0.2680&0.2766\\
	&se$_{w}$ $\hat{\beta}_{iter}$&0.2850&0.2859&0.2919&0.3028&0.3140&0.3316\\
	&se$_{dc}$ $\hat{\beta}_{iter}$&0.3069&0.3086&0.3176&0.3340&0.3500&0.3742\\
	\midrule
	$n=100$&$\hat{\beta}_{1}$&1.0411&1.0408&1.0413&1.0411&1.0402&1.0412\\
	&sd $\hat{\beta}_{1}$&0.2326&0.2315&0.2337&0.2373&0.2420&0.2477\\
	&se $\hat{\beta}_{1}$&0.2212&0.2212&0.2218&0.2229&0.2240&0.2259\\
	&se$_{dc}$ $\hat{\beta}_{1}$&0.2354&0.2359&0.2380&0.2414&0.2458&0.2519\\
	\\
	&$\hat{\beta}_{2}$&1.0353&1.0238&1.0133&1.0041&0.9940&0.9860\\
	&sd $\hat{\beta}_{2}$&0.2153&0.2138&0.2187&0.2239&0.2316&0.2400\\
	&se $\hat{\beta}_{2}$&0.1956&0.1957&0.1964&0.1977&0.1991&0.2010\\
	&se$_{w}$ $\hat{\beta}_{2}$&0.2089&0.2087&0.2099&0.2130&0.2169&0.2221\\
	&se$_{dc}$ $\hat{\beta}_{2}$&0.2135&0.2143&0.2179&0.2239&0.2315&0.2408\\
	\\
	&$\hat{\beta}_{iter}$&1.0386&1.0260&1.0145&1.0044&0.9931&0.9836\\
	&sd $\hat{\beta}_{iter}$&0.2143&0.2126&0.2175&0.2228&0.2311&0.2398\\
	&se $\hat{\beta}_{iter}$&0.1946&0.1958&0.1978&0.2000&0.2024&0.2053\\
	&se$_{w}$ $\hat{\beta}_{iter}$&0.2073&0.2079&0.2101&0.2140&0.2187&0.2248\\
	&se$_{dc}$ $\hat{\beta}_{iter}$&0.2123&0.2129&0.2164&0.2226&0.2298&0.2392\\
	\midrule
	$n=500$&$\hat{\beta}_{1}$&1.0081&1.0080&1.0085&1.0080&1.0082&1.0085\\
	&sd $\hat{\beta}_{1}$&0.1044&0.1048&0.1047&0.1050&0.1056&0.1061\\
	&se $\hat{\beta}_{1}$&0.1035&0.1036&0.1036&0.1038&0.1037&0.1038\\
	&se$_{dc}$ $\hat{\beta}_{1}$&0.1048&0.1049&0.1050&0.1055&0.1057&0.1062\\
	\\
	&$\hat{\beta}_{2}$&1.0066&1.0005&0.9949&0.9885&0.9828&0.9778\\
	&sd $\hat{\beta}_{2}$&0.0962&0.0966&0.0969&0.0970&0.0981&0.0989\\
	&se $\hat{\beta}_{2}$&0.0946&0.0946&0.0946&0.0948&0.0948&0.0949\\
	&se$_{w}$ $\hat{\beta}_{2}$&0.0958&0.0957&0.0956&0.0958&0.0958&0.0962\\
	&se$_{dc}$ $\hat{\beta}_{2}$&0.0955&0.0956&0.0958&0.0964&0.0970&0.0979\\
	\\
	&$\hat{\beta}_{iter}$&1.0074&1.0012&0.9955&0.9891&0.9833&0.9782\\
	&sd $\hat{\beta}_{iter}$&0.0960&0.0964&0.0966&0.0968&0.0977&0.0985\\
	&se $\hat{\beta}_{iter}$&0.0945&0.0949&0.0951&0.0956&0.0959&0.0962\\
	&se$_{w}$ $\hat{\beta}_{iter}$&0.0957&0.0958&0.0959&0.0964&0.0966&0.0972\\
	&se$_{dc}$ $\hat{\beta}_{iter}$&0.0954&0.0954&0.0956&0.0962&0.0967&0.0975\\
		\bottomrule
	\end{tabular}
	\caption{Monte Carlo Results for Linear IV: $n=50, 100, 500$} \label{simulation1_Table}
\end{table}

\subsection{Linear Dynamic Panel Model}

\subsubsection{Random Coefficient}
\label{subsection:simulation-2}
We next explore the finite sample performance of the doubly corrected standard error in the presence of heterogeneous effects (random coefficient) in dynamic panel model. We consider the AR(1) dynamic panel model of Blundell and Bond (1998). For $i=1,...,N$ and $t=1,...,T$,
\begin{eqnarray}
y_{it} &=& \rho_{0} y_{i,t-1} + \eta_{i} + \nu_{it},
\label{simulation2_model}
\end{eqnarray}
where $\eta_{i}$ is an unobserved individual-specific effect and $\nu_{it}$ is an error term. The parameter of interest $\rho_{0}$ is estimated by the difference GMM based on a set of moment conditions: 
\begin{equation}
E[y_{i,t-s}(\Delta y_{it}-\rho_{0}\Delta y_{i,t-1})] = 0, \hspace{0.5em}t=3,...T, \text{ and }s\geq 2,\label{simulation2_moment}
\end{equation}
The moment conditions are derived from taking differences of \eqref{simulation2_model}, and uses the lagged values of $y_{it}$ as instruments. The number of moment conditions is $(T-1)(T-2)/2$.  

The moment conditions are correctly specified if there is a unique parameter that satisfies \eqref{simulation2_moment}. A sufficient condition for this to hold is that the model \eqref{simulation2_model} coincides with the true DGP, but this is unlikely to be true. A reasonable deviation from the assumed model \eqref{simulation2_model} is heterogeneity in $\rho_{0}$ across $i$. We assume the following DGP. For $i=1,...,N$ and $t=1,...T$,
\begin{eqnarray}
\nonumber && y_{it} = \rho_{i} y_{i,t-1} + \eta_{i} + \nu_{it},\\
\nonumber && \eta_{i}\sim N(0,1);~\rho_{i}\sim \Phi(\alpha_{0}\eta_{i});~\nu_{it}\sim N(0,0.5^2),\\
\nonumber && y_{i1} = \frac{\eta_{i}}{1-\rho_{i}}+u_{i1};~u_{i1}\sim N\left(0,\frac{1}{1-\rho_{i}^{2}}\right),
\end{eqnarray} 
where $\Phi(z)$ is the standard normal cdf. At $\alpha_{0}=0$, the model is correctly
specified and $\rho_{i}=\rho_{0}=0.5$. For $\alpha_{0}\neq0$, the effective moment condition model can be written as
\begin{align*}
E[y_{i,t-s}(\Delta y_{it}-\rho\Delta y_{i,t-1})] & = E[y_{i,t-s}(\Delta \nu_{it}+(\rho_{i}-\rho)\Delta y_{i,t-1})]\\
& = E[\rho_{i}y_{i,t-s}\Delta y_{i,t-1}]-\rho (\gamma_{s-1}-\gamma_{s-2})
\end{align*}
where $\gamma_{j}$ is the $j$th autocovariance. The last equation becomes zero at $\rho=E[\rho_{i}]$ if $\rho_{i}$ is independent of the $\{y_{it}\}$ process. If this is the case, then the moment condition model is correctly specified and the estimand is $E[\rho_{i}]$. Otherwise in general, the moment condition model fails to hold at a single unique parameter value because each of the moment condition imposes a restriction
\[\rho = \frac{E[\rho_{i}y_{i,t-s}\Delta y_{i,t-1}]}{\gamma_{s-1}-\gamma_{s-2}}\]
but there is no reason that this should hold at a unique $\rho$ for $s=2,3,...,t-1$. In the DGP, $\eta_{i}$ and $\rho_{i}$ are dependent through $\alpha_{0}$ and a larger $\alpha_{0}$ leads to larger heterogeneity. We vary $\alpha_{0}$ from 0 to 0.3 in steps of 0.05. The pseudo-true value would depend on the instrument set and the value of $\alpha_{0}$ under global misspecification. However, by varying $\alpha_{0}$ by a small amount we try to capture local behavior of the standard errors when the pseudo-true value is close to the true value. The sample sizes are $N=100, 500$ and $T=4, 6$.

We report the simulation results in Tables \ref{simulation2_Table_n100} and \ref{simulation2_Table_n500}, which  are qualitatively similar to the IV setup. Tables \ref{simulation2_Table_n100} and \ref{simulation2_Table_n500} show that the doubly corrected standard errors approximate the standard deviation of the GMM estimators well regardless of misspecification. For the two-step and iterated GMM estimators, the doubly corrected standard errors are as accurate as the Windmeijer correction for small values of $\alpha_{0}$ (including correct specification $\alpha_{0} = 0$) but dominate the other in terms of accuracy for larger values of $\alpha_{0}$. The doubly corrected standard error for the one-step GMM is slightly upward biased for small values of $\alpha_{0}$, but this bias decreases with a larger sample size $N =500$.

	\begin{table}[ptb]
		\centering
		\begin{tabular}{clccccccc}
			\toprule
			&$\alpha_{0}$ &0&0.05&0.1&0.15& 0.2 & 0.25 & 0.3\\
			\midrule
			$N=100$&$\hat{\rho}_{1}$&0.4256&0.4356&0.4571&0.4925&0.5397&0.5926&0.6473\\
			$T=4$&sd $\hat{\rho}_{1}$&0.3324&0.3376&0.3268&0.3173&0.3107&0.3154&0.2763\\
			&se $\hat{\rho}_{1}$&0.3211&0.3215&0.3130&0.3023&0.2899&0.2775&0.2544\\
			&se$_{dc}$ $\hat{\rho}_{1}$&0.3515&0.3520&0.3431&0.3327&0.3196&0.3081&0.2815\\
			\\
			&$\hat{\rho}_{2}$&0.4256&0.4350&0.4554&0.4894&0.5351&0.5876&0.6420\\
			&sd $\hat{\rho}_{2}$&0.3502&0.3552&0.3443&0.3362&0.3327&0.3259&0.2966\\
			&se $\hat{\rho}_{2}$&0.3113&0.3115&0.3032&0.2929&0.2806&0.2680&0.2466\\
			&se$_{w}$ $\hat{\rho}_{2}$&0.3376&0.3377&0.3300&0.3208&0.3087&0.2961&0.2734\\
			&se$_{dc}$ $\hat{\rho}_{2}$&0.3684&0.3673&0.3594&0.3521&0.3375&0.3258&0.3027\\
			\\
			&$\hat{\rho}$&0.4182&0.4276&0.4467&0.4790&0.5234&0.5736&0.6267\\
			&sd $\hat{\rho}$&0.3656&0.3702&0.3619&0.3574&0.3568&0.3551&0.3324\\
			&se $\hat{\rho}$&0.3123&0.3122&0.3038&0.2938&0.2819&0.2688&0.2482\\
			&se$_{w}$ $\hat{\rho}$&0.3483&0.3489&0.3427&0.3350&0.3243&0.3133&0.2909\\
			&se$_{dc}$ $\hat{\rho}$&0.3756&0.3773&0.3695&0.3619&0.3495&0.3398&0.3122\\
			\midrule
			$N=100$&$\hat{\rho}_{1}$&0.4234&0.4272&0.4398&0.4626&0.4992&0.5460&0.6008\\
			$T=6$&sd $\hat{\rho}_{1}$&0.1469&0.1471&0.1468&0.1480&0.1493&0.1503&0.1477\\
			&se $\hat{\rho}_{1}$&0.1458&0.1455&0.1441&0.1418&0.1377&0.1308&0.1221\\
			&se$_{dc}$ $\hat{\rho}_{1}$&0.1537&0.1540&0.1542&0.1546&0.1545&0.1509&0.1441\\
			\\
			&$\hat{\rho}_{2}$&0.4217&0.4249&0.4363&0.4570&0.4909&0.5351&0.5891\\
			&sd $\hat{\rho}_{2}$&0.1630&0.1640&0.1650&0.1667&0.1704&0.1727&0.1708\\
			&se $\hat{\rho}_{2}$&0.1327&0.1324&0.1310&0.1284&0.1242&0.1175&0.1094\\
			&se$_{w}$ $\hat{\rho}_{2}$&0.1635&0.1634&0.1631&0.1626&0.1611&0.1567&0.1493\\
			&se$_{dc}$ $\hat{\rho}_{2}$&0.1628&0.1634&0.1646&0.1668&0.1693&0.1687&0.1643\\
			\\
			&$\hat{\rho}$&0.4167&0.4194&0.4294&0.4476&0.4762&0.5137&0.5606\\
			&sd $\hat{\rho}$&0.1782&0.1799&0.1831&0.1872&0.1972&0.2065&0.2127\\
			&se $\hat{\rho}$&0.1328&0.1325&0.1312&0.1289&0.1250&0.1191&0.1119\\
			&se$_{w}$ $\hat{\rho}$&0.1778&0.1783&0.1794&0.1822&0.1858&0.1887&0.1887\\
			&se$_{dc}$ $\hat{\rho}$&0.1773&0.1784&0.1806&0.1855&0.1916&0.1965&0.1975\\
			\bottomrule
		\end{tabular}
		\caption{Monte Carlo Results for Linear Dynamic Panel: $N=100$ and $T=4, 6$} \label{simulation2_Table_n100}
	\end{table}

	\begin{table}[ptb]
		\centering
		\begin{tabular}{clccccccc}
			\toprule	
			&$\alpha_{0}$ &0&0.05&0.1&0.15& 0.2 & 0.25 & 0.3\\
			\midrule
			$N=500$&$\hat{\rho}_{1}$&0.4879&0.4938&0.5170&0.5531&0.5990&0.6502&0.7016\\
			$T=4$&sd $\hat{\rho}_{1}$&0.1379&0.1363&0.1330&0.1284&0.1220&0.1131&0.1044\\
			&se $\hat{\rho}_{1}$&0.1369&0.1356&0.1322&0.1265&0.1192&0.1107&0.1020\\
			&se$_{dc}$ $\hat{\rho}_{1}$&0.1389&0.1377&0.1347&0.1292&0.1222&0.1136&0.1047\\
			\\
			&$\hat{\rho}_{2}$&0.4893&0.4950&0.5180&0.5539&0.6001&0.6517&0.7034\\
			&sd $\hat{\rho}_{2}$&0.1400&0.1385&0.1357&0.1314&0.1254&0.1164&0.1075\\
			&se $\hat{\rho}_{2}$&0.1361&0.1348&0.1315&0.1258&0.1186&0.1101&0.1014\\
			&se$_{w}$ $\hat{\rho}_{2}$&0.1389&0.1377&0.1349&0.1295&0.1225&0.1138&0.1046\\
			&se$_{dc}$ $\hat{\rho}_{2}$&0.1404&0.1393&0.1368&0.1318&0.1251&0.1165&0.1072\\
			\\
			&$\hat{\rho}$&0.4891&0.4948&0.5178&0.5537&0.5999&0.6515&0.7032\\
			&sd $\hat{\rho}$&0.1403&0.1389&0.1362&0.1319&0.1260&0.1170&0.1080\\
			&se $\hat{\rho}$&0.1362&0.1349&0.1315&0.1259&0.1187&0.1102&0.1016\\
			&se$_{w}$ $\hat{\rho}$&0.1393&0.1382&0.1354&0.1301&0.1232&0.1145&0.1052\\
			&se$_{dc}$ $\hat{\rho}$&0.1408&0.1397&0.1373&0.1323&0.1256&0.1170&0.1077\\
			\midrule
			$N=500$&$\hat{\rho}_{1}$&0.4835&0.4869&0.4997&0.5237&0.5621&0.6130&0.6687\\
			$T=6$&sd $\hat{\rho}_{1}$&0.0691&0.0691&0.0689&0.0681&0.0676&0.0654&0.0615\\
			&se $\hat{\rho}_{1}$&0.0690&0.0688&0.0680&0.0664&0.0637&0.0595&0.0544\\
			&se$_{dc}$ $\hat{\rho}_{1}$&0.0698&0.0697&0.0695&0.0691&0.0681&0.0656&0.0611\\
			\\
			&$\hat{\rho}_{2}$&0.4842&0.4875&0.4998&0.5227&0.5595&0.6089&0.6639\\
			&sd $\hat{\rho}_{2}$&0.0712&0.0714&0.0718&0.0723&0.0735&0.0726&0.0689\\
			&se $\hat{\rho}_{2}$&0.0677&0.0675&0.0667&0.0651&0.0623&0.0581&0.0530\\
			&se$_{w}$ $\hat{\rho}_{2}$&0.0711&0.0711&0.0710&0.0708&0.0701&0.0678&0.0634\\
			&se$_{dc}$ $\hat{\rho}_{2}$&0.0708&0.0709&0.0713&0.0722&0.0730&0.0720&0.0682\\
			\\
			&$\hat{\rho}$&0.4841&0.4874&0.4996&0.5223&0.5585&0.6066&0.6603\\
			&sd $\hat{\rho}$&0.0715&0.0718&0.0723&0.0732&0.0752&0.0757&0.0737\\
			&se $\hat{\rho}$&0.0677&0.0675&0.0667&0.0651&0.0624&0.0583&0.0534\\
			&se$_{w}$ $\hat{\rho}$&0.0715&0.0715&0.0715&0.0717&0.0720&0.0713&0.0685\\
			&se$_{dc}$ $\hat{\rho}$&0.0712&0.0713&0.0718&0.0730&0.0746&0.0749&0.0725\\
			\bottomrule
		\end{tabular}
		\caption{Monte Carlo Results for Linear Dynamic Panel: $N=500$ and $T=4, 6$} \label{simulation2_Table_n500}
	\end{table}

\subsubsection{Misspecified Lag Length}
\label{subsection:simulation-3}
We use the baseline linear panel model of Windmeijer (2005) allowing for possible lag length misspecification. The model is 
\begin{equation}
y_{it}  =\beta_{0} x_{it}  + \eta_{i}+v_{it},
\label{simulation3_model}
\end{equation}
for $i = 1, ..., N$ and $t = 1, ..., T$. The unknown parameter of interest is $\beta_{0}$, and the
regressor $x_{it}$ is predetermined with respect
to $v_{it}$, i.e., $E(x_{it}v_{it+s})=0$ for $s=0,...,T-t.$ We use the first differenced GMM estimator and the number of moment conditions is $T(T-1)/2$ as in Section \ref{Section_Examples_Panel}.

The DGP is
\begin{align}
\label{simulation3_dgp}y_{it}  & =\beta_{0}x_{it}+\alpha_{0} x_{it-1}+\eta_{i}+v_{it},\\ \nonumber
x_{it}  & =0.5x_{it-1}+\eta_{i}+0.5v_{it-1}+\epsilon_{it},\\ \nonumber
\eta_{i} & \sim N(0,1),~~\epsilon_{it}\sim N(0,1),\\ \nonumber
v_{it}  & =\delta_{i}\tau_{t}\omega_{it},~~\omega_{it}\sim\chi
_{1}^{2}-1,\\ \nonumber 
\delta_i &\sim \textnormal{Uniform}[0.5, 1.5],~~\tau_t = 0.5 + 0.1 (t-1). 
\end{align}
We generate initial $50$ time periods with $\tau_{t}=0.5$ for $t=-49,\ldots,0$ and $x_{i, -49}\sim N(\eta_{i}/0.5,1/0.75)$ same as Windmeijer (2005). The parameter $\alpha_{0}$ in \eqref{simulation3_dgp} governs the degree of misspecification. When  $\alpha_{0}=0$, the model \eqref{simulation3_model} is correctly specified which reduces to that of Windmeijer (2005).

The model \eqref{simulation3_model} is misspecified for $\alpha_{0}\neq0$. We discuss the pseudo-true value and its interpretation. Since $0=E[x_{is}\Delta v_{it}]$, the effective moment condition can be written as
\begin{equation}
\label{ex3-mc}
0=E[x_{is}(\Delta y_{it}-\beta^{*}\Delta x_{it})] = (\beta_{0}-\beta^{*})E[x_{is}\Delta x_{it}] + \alpha_{0}E[x_{is}\Delta x_{it-1}]
\end{equation}
for all $1\leq s<t$ and $2\leq t\leq T$. 

First, consider $T=2$ where there is only one moment condition so that the model is just-identified. By setting $s=1$ and $t=2$, the unique solution to \eqref{ex3-mc} is $\beta^{*}=\beta_{0}-\alpha_{0}$. Since \eqref{ex3-mc} equals to zero at  $\beta^{*}$, the moment condition is not misspecified although the lag is misspecified. Since the model is just-identified $\beta^{*}=\beta_{0}-\alpha_{0}$ is considered as the true value but this is not equal to $\beta_{0}$ unless $\alpha_{0}=0$. An implication is that model misspecification and pseudo-true values are not only pertinent to over-identified models. 

For $T\geq3$, the model is overidentified and each of the moment condition holds at
\begin{equation}
\beta^{*}_{t,s}=\beta_{0} + \alpha_{0} \frac{E[x_{i,t-1}x_{is}]-E[x_{i,t-2}x_{is}]}{E[x_{it}x_{is}]-E[x_{i,t-1}x_{is}]}.
\end{equation}
Note that $\beta^{*}_{t,s}$ can vary across $t$ and $s$ for a nonzero $\alpha_{0}$. For example, $\beta^{*}_{3,1} = \beta_{0} + 2\alpha_{0}$ and $\beta^{*}_{3,2} = \beta_{0}-\alpha_{0}$. The pseudo-true value $\beta^{*}$ is a weighted average of $\beta_{t,s}^{*}$'s given $T$ and the weight matrix. One can interpret that $\beta^{*}$ represents the causal effects of past and present values of $x_{it}$'s on $y_{it}$ in the true DGP.


Tables \ref{simulation3_Table_n100} and \ref{simulation3_Table_n500} report estimation results for $\beta_{0}=1,$ $N=100, 500$ and $T=4,6$. The degree of misspecification $\alpha_{0}$ is varied across $\{ 0, 0.05, 0.1, 0.2, 0.3 \}$. The first column $(\alpha_{0} =0)$ in Table \ref{simulation3_Table_n100} replicates Monte Carlo studies in Windmeijer (2005, Table 1). Tables \ref{simulation3_Table_n100} and \ref{simulation3_Table_n500} also report the rejection rates of the $J$ test with the nominal size of 5\%. Although the rejection rate increases as the degree of misspecification increases and the sample size increase, the $J$ test tends to not correctly detect the violation of the overidentifying moment restrictions in small samples.

The implication of the results in Tables \ref{simulation3_Table_n100} and \ref{simulation3_Table_n500} are largely unchanged as in two previous simulation experiments; doubly corrected standard errors approximate the standard deviations well regardless of model misspecification. In this simulation experiment, the Windmeijer correction works best under correct specification but becomes downward biased as $\alpha_{0}$ increases. Note that deviation from the correct specification makes the bias of the conventional standard error and the Windmeijer corrected standard error larger, and this bias does not disappear with a larger sample size of $N=500$.

\begin{table}[ptb]
	\centering
	\begin{tabular}{clccccc}
		\toprule
		&$\alpha_{0}$&0 &0.05&0.1&0.2&0.3\\
		\midrule
		$N=100$&$\hat{\beta}_{1}$&0.9793&0.9268&0.8744&0.7702&0.6647\\
		$T = 4$&sd $\hat{\beta}_{1}$&0.1521&0.1538&0.1588&0.1727&0.1925\\
		&se $\hat{\beta}_{1}$&0.1469&0.1465&0.1472&0.1513&0.1589\\
		&se$_{dc}$ $\hat{\beta}_{1}$&0.1546&0.1563&0.1604&0.1744&0.1944\\
		\cmidrule{2-7}
		& $J$ test &0.0779&0.0836&0.1127&0.2899&0.6127\\
		\cmidrule{2-7}
		\\
		&$\hat{\beta}_{2}$&0.9849&0.9322&0.8773&0.7587&0.6238\\
		&sd $\hat{\beta}_{2}$&0.1404&0.1451&0.1552&0.1843&0.2207\\
		&se $\hat{\beta}_{2}$&0.1243&0.1242&0.1253&0.1303&0.1381\\
		&se$_{w}$ $\hat{\beta}_{2}$&0.1390&0.1415&0.1473&0.1669&0.1919\\
		&se$_{dc}$ $\hat{\beta}_{2}$&0.1343&0.1390&0.1482&0.1775&0.2146\\
		\cmidrule{2-7}
		& $J$ test &0.0301&0.0391&0.0767&0.2605&0.5404 \\
		\cmidrule{2-7}
		\\
		&$\hat{\beta}$&0.9858&0.9334&0.8781&0.7533&0.5977\\
		&sd $\hat{\beta}$&0.1417&0.1474&0.1600&0.2000&0.2524\\
		&se $\hat{\beta}$&0.1243&0.1242&0.1253&0.1303&0.1381\\
		&se$_{w}$ $\hat{\beta}$&0.1393&0.1426&0.1507&0.1806&0.2230\\
		&se$_{dc}$ $\hat{\beta}$&0.1352&0.1406&0.1517&0.1896&0.2391\\
		\cmidrule{2-7}
		& $J$ test &0.0295&0.0377&0.0727&0.2429&0.4983\\
		\midrule
		$N=100$&$\hat{\beta}_{1}$&0.9755&0.9411&0.9060&0.8368&0.7676\\
		$T=6$&sd $\hat{\beta}_{1}$&0.1027&0.1051&0.1083&0.1172&0.1288\\
		&se $\hat{\beta}_{1}$&0.1002&0.1004&0.1013&0.1037&0.1077\\
		&se$_{dc}$ $\hat{\beta}_{1}$&0.1056&0.1075&0.1107&0.1192&0.1306\\
		\cmidrule{2-7}
		& $J$ test &0.3465&0.3533&0.4157&0.6786&0.9239\\
		\cmidrule{2-7}
		\\
		&$\hat{\beta}_{2}$&0.9833&0.9466&0.9080&0.8238&0.7318\\
		&sd $\hat{\beta}_{2}$&0.0906&0.0948&0.1017&0.1213&0.1431\\
		&se $\hat{\beta}_{2}$&0.0716&0.0720&0.0731&0.0760&0.0801\\
		&se$_{w}$ $\hat{\beta}_{2}$&0.0905&0.0930&0.0978&0.1117&0.1285\\
		&se$_{dc}$ $\hat{\beta}_{2}$&0.0836&0.0876&0.0944&0.1131&0.1357\\
		\cmidrule{2-7}
		& $J$ test&0.0205&0.0277&0.0582&0.2762&0.6453\\
		\cmidrule{2-7}
		\\
		&$\hat{\beta}$&0.9857&0.9484&0.9083&0.8124&0.6885\\
		&sd $\hat{\beta}$&0.0946&0.0998&0.1099&0.1427&0.1858\\
		&se $\hat{\beta}$&0.0716&0.0720&0.0731&0.0760&0.0801\\
		&se$_{w}$ $\hat{\beta}$&0.0937&0.0976&0.1054&0.1320&0.1722\\
		&se$_{dc}$ $\hat{\beta}$&0.0866&0.0916&0.1006&0.1287&0.1675\\
		\cmidrule{2-7}
		& $J$ test&0.0199&0.0259&0.0551&0.2553&0.5980\\
		\bottomrule
	\end{tabular}
	\caption{Monte Carlo Results for Linear Panel Model: $N=100$ and $T=4, 6$}\label{simulation3_Table_n100}
\end{table}

\begin{table}[ptb]
	\centering
	\begin{tabular}{clccccc}
		\toprule
		&$\alpha_{0}$&0&0.05&0.1&0.2&0.3\\
		\midrule
		$N=500$&$\hat{\beta}_{1}$&0.9958&0.9406&0.8853&0.7754&0.6650\\
		$T=4$&sd $\hat{\beta}_{1}$&0.0685&0.0695&0.0716&0.0787&0.0889\\
		&se $\hat{\beta}_{1}$&0.0679&0.0677&0.0679&0.0698&0.0734\\
		&se$_{dc}$ $\hat{\beta}_{1}$&0.0686&0.0696&0.0716&0.0787&0.0889\\
		\cmidrule{2-7}
		& $J$ test&0.0732&0.1212&0.3934 &0.9524 &1.0000 \\
		\cmidrule{2-7}
		\\
		&$\hat{\beta}_{2}$&0.9970&0.9443&0.8892&0.7660&0.6225\\
		&sd $\hat{\beta}_{2}$&0.0652&0.0676&0.0730&0.0891&0.1092\\
		&se $\hat{\beta}_{2}$&0.0632&0.0630&0.0634&0.0657&0.0693\\
		&se$_{w}$ $\hat{\beta}_{2}$&0.0648&0.0662&0.0694&0.0798&0.0930\\
		&se$_{dc}$ $\hat{\beta}_{2}$&0.0634&0.0658&0.0708&0.0865&0.1060\\
		\cmidrule{2-7}
		& $J$ test&0.0447&0.1014&0.3824&0.9426&0.9996\\
		\cmidrule{2-7}
		\\
		&$\hat{\beta}$&0.9970&0.9446&0.8897&0.7635&0.6014\\
		&sd $\hat{\beta}$&0.0652&0.0678&0.0738&0.0946&0.1238\\
		&se $\hat{\beta}$&0.0632&0.0630&0.0634&0.0657&0.0693\\
		&se$_{w}$ $\hat{\beta}$&0.0648&0.0662&0.0698&0.0839&0.1052\\
		&se$_{dc}$ $\hat{\beta}$&0.0634&0.0659&0.0715&0.0911&0.1185\\
		\cmidrule{2-7}
		&$J$ test &0.0445&0.1004&0.3822&0.9410&0.9995\\
		\midrule
		$N=500$&$\hat{\beta}_{1}$&0.9947&0.9564&0.9181&0.8423&0.7662\\
		$T=6$&sd $\hat{\beta}_{1}$&0.0473&0.0482&0.0498&0.0543&0.0600\\
		&se $\hat{\beta}_{1}$&0.0469&0.0470&0.0473&0.0485&0.0504\\
		&se$_{dc}$ $\hat{\beta}_{1}$&0.0475&0.0484&0.0499&0.0543&0.0602\\
		\cmidrule{2-7}
		&$J$ test &0.3557&0.4796&0.7820&0.9995&1.0000\\
		\cmidrule{2-7}
		\\
		&$\hat{\beta}_{2}$&0.9968&0.9582&0.9175&0.8276&0.7251\\
		&sd $\hat{\beta}_{2}$&0.0432&0.0454&0.0494&0.0602&0.0725\\
		&se $\hat{\beta}_{2}$&0.0408&0.0410&0.0414&0.0428&0.0448\\
		&se$_{w}$ $\hat{\beta}_{2}$&0.0431&0.0445&0.0471&0.0548&0.0640\\
		&se$_{dc}$ $\hat{\beta}_{2}$&0.0413&0.0434&0.0470&0.0573&0.0694\\
		\cmidrule{2-7}
		&$J$ test &0.0388&0.1139&0.4500&0.9906&1.0000\\
		\cmidrule{2-7}
		\\
		&$\hat{\beta}$&0.9969&0.9584&0.9174&0.8219&0.6961\\
		&sd $\hat{\beta}$&0.0433&0.0457&0.0504&0.0659&0.0882\\
		&se $\hat{\beta}$&0.0408&0.0410&0.0414&0.0428&0.0448\\
		&se$_{w}$ $\hat{\beta}$&0.0431&0.0447&0.0479&0.0592&0.0770\\
		&se$_{dc}$ $\hat{\beta}$&0.0414&0.0437&0.0479&0.0620&0.0823\\
		\cmidrule{2-7}
		&$J$ test &0.0387&0.1131&0.4489&0.9902&1.0000\\
		\bottomrule
	\end{tabular}
	\caption{Monte Carlo Results for Linear Panel Model: $N=500$ and $T=4, 6$}\label{simulation3_Table_n500}
\end{table}

\subsection{Size}

This section investigates the small sample performance of the $t$ tests with the proposed standard errors. 
We report the size of the $t$ tests under the correct specification for the simulation setups considered in previous sections. 

Based on the one-step, two-step, and iterated GMM estimators, Table \ref{simulation_size} evaluates the size of the $t$ tests for nominal size 5\%  using various standard errors. In the column labeled $t$, we report the size of the test based on conventional heteroskedasticity-robust standard errors. $t_w$ is based on the Windmeijer corrected standard errors while $t_{dc}$ is based on the doubly corrected standard errors in this paper. Further, we report the size of the bootstrap $t$ test using the doubly corrected standard errors ($t_{dc}$-$bs$) which is equivalent to the misspecification-robust (MR) bootstrap of Lee (2014). To calculate bootstrap critical values, 1000 additional bootstrap replications are performed per Monte Carlo replication. Results for larger sample sizes ($n=500$ in Table \ref{simulation1_Table}, Table \ref{simulation2_Table_n500} and \ref{simulation3_Table_n500}) are not reported here for brevity as the results are similar across different methods. 

For Setups 1 and 2, tests using the conventional standard errors ($t$) are severely oversized. Using the Windmeijer corrected standard errors ($t_w$), and the doubly corrected standard errors ($t_{dc}$) for the two-step and iterated estimator improve the size and perform similarly, although both are moderately oversized. The size distortion decreases when we increase the sample size. Using the MR bootstrap with the doubly corrected standard errors ($t_{dc}$-$bs$) improves the size of the test dramatically. The excellent size property of the MR bootstrap $t$ test is theoretically justified by its asymptotic refinements, which is formally shown by Lee (2014). Perhaps surprisingly, using the doubly corrected standard errors for the $t$ test and bootstrap $t$ test improves size properties considerably for the one-step estimator where the Windmeijer correction is not available. 

For Setup 3, $t$ based on the one-step estimator and $t_{w}$ with the two-step estimator have good size properties, and the same results can be found in Windmeijer (2005, Fig 1.). $t_{dc}$ has similar size properties to $t_{w}$, but is slightly undersized for the one-step estimator and slightly oversized for the two-step and iterated estimators. The MR bootstrap test is slightly undersized. Windmeijer (2005) and Bond and Windmeijer (2005) report similar results for the nonparametric bootstrap test of Hall and Horowitz (1996) based on the two-step estimator and explain that the performance of the bootstrap deteriorates with an increasing number of moment conditions. Since $t_{dc}$ becomes oversized with the number of moment conditions, this suggests that some components of the doubly corrected variance estimator may be sensitive to the number of moment conditions. This issue deserves further future investigation.

\renewcommand{\tabcolsep}{3pt}
\begin{table}[ptb]
\centering
\begin{threeparttable}
    \begin{tabular}{ll ccccc ccccc ccc}
    \toprule
    &   & \multicolumn{3}{c}{One-step}  &  &  \multicolumn{4}{c}{Two-step} & & \multicolumn{4}{c}{Iterated}  \\
      \cmidrule{3-5}  \cmidrule{7-10} \cmidrule{12-15} 
         & &  $t$ & $t_{dc}$  & $t_{dc}$-$bs$&&  $t$ & $t_{w}$ & $t_{dc}$ & $t_{dc}$-$bs$&& $t$ & $t_{w}$ & $t_{dc}$ & $t_{dc}$-$bs$ \\
 \midrule      
    \multicolumn{9}{l}{Setup 1: Cross-sectional IV (Section \ref{subsection:simulation-1}) }\\ \\
$n=50$&   &  0.109 & 0.084  & 0.072&&  0.142 & 0.112 & 0.110 & 0.077&& 0.151 & 0.123 & 0.122 & 0.071 \\ \\
 $n=100$&   &  0.079 & 0.065  & 0.064&&  0.096 & 0.082 & 0.082 & 0.069 && 0.100 & 0.085 & 0.085 & 0.068 \\
    \midrule
       \multicolumn{9}{l}{Setup 2: Dynamic Panel (Section \ref{subsection:simulation-2})}\\  \\
$N=100, T=4$ &  &  0.074 & 0.067  & 0.060 &&  0.096 & 0.076 & 0.075 & 0.061&& 0.106 & 0.080 & 0.079 & 0.057 \\ \\
              $N=100, T=6$  &&  0.089 & 0.082  & 0.069&&  0.154 & 0.082 & 0.091 & 0.065&& 0.178 & 0.078 & 0.090 & 0.055 \\
   \midrule
          \multicolumn{9}{l}{Setup 3: Dynamic Panel (Section \ref{subsection:simulation-3})}\\  \\
 $N=100, T=4$&   &  0.047 & 0.039 & 0.030 &&  0.073 & 0.047 & 0.056 & 0.027 && 0.078 & 0.049 & 0.058 & 0.025 \\ \\
              $N=100, T=6$  & &  0.060 & 0.046  & 0.041&&  0.123 & 0.055& 0.074 & 0.034&& 0.137 & 0.057 & 0.076 &0.025 \\
          \bottomrule
    \end{tabular}
       \caption{Finite Sample Sizes of the $t$-test (Nominal Size 5\%)}
    \label{simulation_size}
  \end{threeparttable}
\end{table}%

%


\appendix
\section*{Appendix A: Proofs}

In the proofs, ``LLN'' refers to the Weak Law of Large Numbers, ``CLT'' refers to the multivariate Lindeberg-L\'{e}vy Central Limit Theorem (e.g. Theorem 6.3. of Hansen, 2020) and ``CMT'' refers to the continuous mapping theorem.


Lemma \ref{lemma_matrix} is on geometric expansion of a matrix. It builds on Corollary 1 of Magdalinos (1992).

\begin{lemma}\label{lemma_matrix} Let $ X_n$ and $Y_n$ be square random matrices. If $X_n^{-1}$ and $(X_{n} + Y_{n}/\sqrt{n})^{-1}$ exist and $X_{n}^{-1}$ and $Y_n$ are of order $O_p(1)$, then following holds for any nonnegative integer $q$,
\begin{equation*}
\left(X_n + \frac{1}{\sqrt{n}} Y_n\right)^{-1} = \sum_{j=0}^{q} \left(-\frac{1}{\sqrt{n}} X_{n}^{-1} Y_n\right)^{j} X_n^{-1} + O_p(n^{-(q+1)/2}).
\end{equation*}
\end{lemma}

\noindent
\textbf{Proof of Lemma \ref{lemma_matrix}: } Let $S_n = X_n^{-1}Y_n$ and consider the following identity,
\[
\sum_{j=0}^{q} \left(-\frac{1}{\sqrt{n}}\right)^j S_n^{j} \left(I + \frac{1}{\sqrt{n}} S_n\right) = I - \left(-\frac{1}{\sqrt{n}}\right)^{q+1} S_{n}^{q+1}. 
\]
Using $I + n^{-1/2} S_n = X_n^{-1} (X_n + n^{-1/2} Y_n)$, $S_n = X_n^{-1}Y_n$ and rearranging terms, we have
\begin{align*}
\left(X_n + \frac{1}{\sqrt{n}} Y_n\right)^{-1} &= \sum_{j=0}^{q} \left(-\frac{1}{\sqrt{n}}\right)^j S_n^{j}  X_n^{-1} + \left(-\frac{1}{\sqrt{n}}\right)^{q+1} S_{n}^{q+1} \left(X_n + \frac{1}{\sqrt{n}} Y_n\right)^{-1}\\
&= \sum_{j=0}^{q} \left(-\frac{1}{\sqrt{n}}\right)^j \left(X_n^{-1}Y_n\right)^{j}  X_n^{-1} + \left(-\frac{1}{\sqrt{n}}\right)^{q+1} O_p(1)
\end{align*}
by the assumptions of the lemma. \qed

\bigskip

Let 
\begin{align*}
\frac{\partial\Omega(\theta)}{\partial\theta_{[j]}}=&E\left[\frac{\partial\Omega_{n}(\theta)}{\partial\theta_{[j]}}\right],\\
\frac{\partial^{2}\Omega(\theta)}{\partial\theta_{[j]}\partial\theta_{[l]}}=&E\left[\frac{\partial^{2}\Omega_{n}(\theta)}{\partial\theta_{[j]}\partial\theta_{[l]}}\right].
\end{align*}
This is a slight abuse of notation because the LHS does not mean differentiation of the expectation. Lemma \ref{lemma_convergence} establishes some useful convergence results.

\begin{lemma}
	Under Assumption 1, the followings hold:
	\begin{align}
	\label{gn112}g_{n}(\theta_{1})-g_{1}&=O_{p}(n^{-1/2}),\\
	\label{gn212}g_{n}(\theta_{2})-g_{2}&=O_{p}(n^{-1/2}),\\
	\label{Gn12}G_{n}-G&=O_{p}(n^{-1/2}),\\
	\label{Wn12}W_{n}-W&=O_{p}(n^{-1/2}),\\
	\label{On12}\Omega_{n}(\theta_{1})-\Omega_{1}&=O_{p}(n^{-1/2}),\\
	\label{iWn12}W_{n}^{-1}-W^{-1}&=O_{p}(n^{-1/2}),\\
	\label{iOn12}[\Omega_{n}(\theta_{1})]^{-1}-[\Omega_{1}]^{-1}&=O_{p}(n^{-1/2}),
	\end{align}
	and for each $j$, $l$th element of $\theta$,
	\begin{align}
	\label{DOn12}\frac{\partial\Omega_{n}(\theta_{1})}{\partial\theta_{[j]}}-\frac{\partial\Omega(\theta_{1})}{\partial\theta_{[j]}}&=O_{p}(n^{-1/2}),\\
	\label{D2On12}\frac{\partial^{2}\Omega_{n}(\theta_{1})}{\partial\theta_{[j]}\partial\theta_{[l]}}-\frac{\partial^{2}\Omega(\theta_{1})}{\partial\theta_{[j]}\partial\theta_{[l]}}&=O_{p}(n^{-1/2}),
	\end{align}
	In addition, if $\hat{\theta}_{1}-\theta_{1} = O_{p}(n^{-1/2})$, then 
	\begin{align}
	\label{Onhat12}\Omega_{n}(\hat{\theta}_{1})-\Omega_{1}&=O_{p}(n^{-1/2}),\\
	\label{iOnhat12}[\Omega_{n}(\hat{\theta}_{1})]^{-1}-[\Omega_{1}]^{-1}&=O_{p}(n^{-1/2}).
	\end{align}
	\label{lemma_convergence}
\end{lemma}

\noindent
\textbf{Proof of Lemma \ref{lemma_convergence}:} 
Under Assumption 1 (ii) and (v), \eqref{gn112}-\eqref{On12} immediately follow by CLT. 

Since $g(X_{i},\theta)$ is linear in $\theta$,
\begin{align}
\frac{\partial\Omega_{n}(\theta_{1})}{\partial\theta_{[j]}} =& \frac{1}{n}\sum_{i=1}^{n}g(X_{i},\theta_{1})\frac{\partial g(X_{i},\theta_{1})'}{\partial\theta_{[j]}} + \frac{1}{n}\sum_{i=1}^{n}\frac{\partial g(X_{i},\theta_{1})}{\partial\theta_{[j]}}g(X_{i},\theta_{1})',\\
\frac{\partial^{2}\Omega_{n}(\theta_{1})}{\partial\theta_{[j]}\partial\theta_{[l]}} =&  \frac{1}{n}\sum_{i=1}^{n}\frac{\partial g(X_{i},\theta_{1})}{\partial\theta_{[l]}}\frac{\partial g(X_{i},\theta_{1})'}{\partial\theta_{[j]}} +\frac{1}{n}\sum_{i=1}^{n}\frac{\partial g(X_{i},\theta_{1})}{\partial\theta_{[j]}} \frac{\partial g(X_{i},\theta_{1})'}{\partial\theta_{[l]}}.
\end{align}
Under Assumption 1 (ii) and (v), we apply CLT and the Cauchy-Schwarz inequality to have \eqref{DOn12} and \eqref{D2On12}.

Next we show \eqref{Onhat12}. Observing that the third derivative of $\Omega_{n}(\theta)$ with respect to $\theta$ equals to zero, we use the second-order Taylor expansion to obtain
\begin{align*}
\Omega_{n}(\hat{\theta}_{1}) =& \Omega_{1} + (\Omega_{n}(\theta_{1})-\Omega_{1}) + \sum_{j}\frac{\partial\Omega(\theta_{1})}{\partial\theta_{[j]}}(\hat{\theta}_{1[j]}-\theta_{1[j]})\\
& + \sum_{j}\left(\frac{\partial\Omega_{n}(\theta_{1})}{\partial\theta_{[j]}}-\frac{\partial\Omega(\theta_{1})}{\partial\theta_{[j]}}\right)(\hat{\theta}_{1[j]}-\theta_{1[j]})\\
&+ \frac{1}{2}\sum_{j}\sum_{l}\frac{\partial^{2}\Omega(\theta_{1})}{\partial\theta_{[j]}\partial\theta_{[l]}}(\hat{\theta}_{1[j]}-\theta_{1[j]})(\hat{\theta}_{1[l]}-\theta_{1[l]})\\
&+ \frac{1}{2}\sum_{j}\sum_{l}\left(\frac{\partial^{2}\Omega_{n}(\theta_{1})}{\partial\theta_{[j]}\partial\theta_{[l]}}-\frac{\partial^{2}\Omega(\theta_{1})}{\partial\theta_{[j]}\partial\theta_{[l]}}\right)(\hat{\theta}_{1[j]}-\theta_{1[j]})(\hat{\theta}_{1[l]}-\theta_{1[l]}).
\end{align*}
By Assumption 1 (ii) and (v), the Cauchy-Schwarz inequality implies
\begin{equation}
\label{finiteDO}
\frac{\partial\Omega(\theta_{1})}{\partial\theta_{[j]}}<\infty~\text{ and }~\frac{\partial^{2}\Omega(\theta_{1})}{\partial\theta_{[j]}\partial\theta_{[l]}}<\infty.
\end{equation}
Provided that $\hat{\theta}_{1}-\theta_{1}=O_{p}(n^{-1/2})$, \eqref{Onhat12} holds by \eqref{On12}, \eqref{DOn12}, \eqref{D2On12}, and \eqref{finiteDO}.

Finally, applying Lemma \ref{lemma_matrix} with $q=0$ to \eqref{Wn12}, \eqref{On12}, and \eqref{Onhat12}, we obtain \eqref{iWn12}, \eqref{iOn12}, and \eqref{iOnhat12}, respectively.\qed

\bigskip

\noindent
\textbf{Proof of Theorem \ref{TMR}:} The proof consists of two parts. We first derive the asymptotic variance matrices of the one-step and two-step GMM estimators without assuming correct specification of the moment condition. Then we show that the doubly corrected variance estimators are consistent. 

First, Lemma \ref{lemma_convergence}, CMT, and the Woodbury matrix identity together imply that
\begin{align}
\notag (G_{n}'W_{n}^{-1}G_{n})^{-1} =& (G'W^{-1}G)^{-1}(I_{k} + (G_{n}'W_{n}^{-1}G_{n}-G'W^{-1}G)(G'W^{-1}G)^{-1})^{-1}\\
\label{gwg}=& (G'W^{-1}G)^{-1}(I_{k}+o_{p}(1))
\end{align}
and similarly
\begin{equation}
\label{gog} (G_{n}'[\Omega_{n}(\theta_{1})]^{-1}G_{n})^{-1} = (G'\Omega_{1}^{-1}G)^{-1}(I_{k}+o_{p}(1)).
\end{equation}

The FOC of the one-step GMM estimator is
\begin{equation}
\label{FOC1}
0 = G_{n}'W_{n}^{-1}g_{n}(\hat{\theta}_{1})
\end{equation}
which holds with probability approaching zero by Assumption 1(i). By expanding \eqref{FOC1} around the pseudo-true value $\theta_{1}$ and rearranging, we have
\begin{align*}
\label{FOC1e}
\hat{\theta}_{1}-\theta_{1} =& -(G_{n}'W_{n}^{-1}G_{n})^{-1}G_{n}'W_{n}^{-1}g_{n}(\theta_{1})\\
=&- (G'W^{-1}G)^{-1}(I_{k}+o_{p}(1))G_{n}'W_{n}^{-1}g_{n}(\theta_{1}).
\end{align*}
By Lemma \ref{lemma_convergence} and the population FOC $G'W^{-1}g_{1}=0$, 
\begin{align*}
G_{n}'W_{n}^{-1}g_{n}(\theta_{1}) = G'W^{-1}g_{n}(\theta_{1}) + G_{n}'W^{-1}g_{1} - G'W^{-1}W_{n}W^{-1}g_{1} + O_{p}(n^{-1}).
\end{align*}
Let 
\begin{equation*}
m_{1i} = G'W^{-1}g(X_{i},\theta_{1}) + G(X_{i})'W^{-1}g_{1} - G'W^{-1}W(X_{i})W^{-1}g_{1}.
\end{equation*}
Since $Em_{1i}=0$, by CLT under Assumption 1(v)
\begin{equation}
\frac{1}{\sqrt{n}}\sum_{i=1}^{n}m_{1i}\xrightarrow{d}N(0,Em_{1i}m_{1i}')
\end{equation}
as $n\rightarrow\infty$. Now we can write
\begin{align}
\notag \sqrt{n}(\hat{\theta}_{1}-\theta_{1}) =&  - (G'W^{-1}G)^{-1}(I_{k}+o_{p}(1))\left(\frac{1}{\sqrt{n}}\sum_{i=1}^{n}m_{1i} + O_{p}(n^{-1/2})\right)\\
\label{gmm1}=&  - (G'W^{-1}G)^{-1}\frac{1}{\sqrt{n}}\sum_{i=1}^{n}m_{1i}+o_{p}(1)\\
\label{t1Op}\xrightarrow{d}&N(0,V_{1})
\end{align}
as $n\rightarrow\infty$ where $V_{1}$ is the asymptotic variance matrix of the one-step GMM under misspecification given by
\begin{equation}
V_{1} = (G'W^{-1}G)^{-1}Em_{1i}m_{1i}' (G'W^{-1}G)^{-1}.
\end{equation}

The FOC of the two-step GMM estimator is
\begin{equation}
\label{FOC2}
0 = G_{n}'[\Omega_{n}(\hat{\theta}_{1})]^{-1}g_{n}(\hat{\theta}_{2}).
\end{equation}
By expanding $g_{n}(\hat{\theta}_{2})$ around $\theta_{2}$ and then expanding $[\Omega_{n}(\hat{\theta}_{1})]^{-1}$ around $\theta_{1}$,
\begin{align}
\notag \hat{\theta}_{2}-\theta_{2} =&  -(G_{n}'[\Omega_{n}(\hat{\theta}_{1})]^{-1}G_{n})^{-1}G_{n}[\Omega_{n}(\hat{\theta}_{1})]^{-1}g_{n}(\theta_{2})\\
\label{misexp}=&-(G_{n}'[\Omega_{n}(\theta_{1})]^{-1}G_{n})^{-1}G_{n}[\Omega_{n}(\theta_{1})]^{-1}g_{n}(\theta_{2}) + D_{n}^{*}(\hat{\theta}_{1}-\theta_{1}) + R_{n}^{*},
\end{align}
where $D_{n}^{*} = F_{1n}^{*} + F_{2n}^{*}$ and $R_{n}^{*}$ is the remainder term.

First, $F_{1n}^{*}=o_{p}(1)$ by Lemma \ref{lemma_convergence} and the population FOC $G'\Omega_{1}^{-1}g_{2}=0$. By Lemma \ref{lemma_convergence}, $F_{2n}^{*}=D^{*}+o_{p}(1)$ where 
\begin{align*}
D^{*}[.,j] &= \left(G'\Omega_{1}^{-1}G\right)^{-1}G'\Omega_{1}^{-1}\left.\frac{\partial\Omega(\theta)}{\partial\theta_{[j]}}\right|_{\theta=\theta_{1}}\Omega_{1}^{-1}g_{2}.
\end{align*}
Hence we can write $D_{n}^{*} = D^{*} + o_{p}(1)$. 

Next we show $R_{n}^{*}=O_{p}(n^{-1})$. Let 
\begin{equation}
Q_{n}(\theta)=\left\{G_{n}'[\Omega_{n}(\theta)]^{-1}G_{n}\right\}^{-1}G_{n}'[\Omega_{n}(\theta)]^{-1}g_{n}(\theta_{2}).
\end{equation}
Since the third derivative of $\Omega_{n}(\theta)$ w.r.t. $\theta$ equals to zero, the fifth derivative of $Q_{n}(\theta)$ w.r.t. $\theta$ equals to zero by the chain rule. Thus,
\begin{align*}
R_{n} =& \frac{1}{2!}\sum_{j}\sum_{l}\frac{\partial^{2}Q_{n}(\theta_{1})}{\partial\theta_{[j]}\partial\theta_{[l]}}(\hat{\theta}_{1[j]}-\theta_{1[j]})(\hat{\theta}_{1[l]}-\theta_{1[l]})\\
& + \frac{1}{3!}\sum_{j}\sum_{l}\sum_{m}\frac{\partial^{3}Q_{n}(\theta_{1})}{\partial\theta_{[j]}\partial\theta_{[l]}\partial\theta_{[m]}}(\hat{\theta}_{1[j]}-\theta_{1[j]})(\hat{\theta}_{1[l]}-\theta_{1[l]})(\hat{\theta}_{1[m]}-\theta_{1[m]})\\
& + \frac{1}{4!}\sum_{j}\sum_{l}\sum_{m}\sum_{s}\frac{\partial^{4}Q_{n}(\theta_{1})}{\partial\theta_{[j]}\partial\theta_{[l]}\partial\theta_{[m]}\partial\theta_{[s]}}(\hat{\theta}_{1[j]}-\theta_{1[j]})(\hat{\theta}_{1[l]}-\theta_{1[l]})(\hat{\theta}_{1[m]}-\theta_{1[m]})(\hat{\theta}_{1[s]}-\theta_{1[s]}).
\end{align*}
By LLN under Assumption 1, $\frac{\partial^{2}Q_{n}(\theta_{1})}{\partial\theta_{[j]}\partial\theta_{[l]}}=O_{p}(1)$, $\frac{\partial^{3}Q_{n}(\theta_{1})}{\partial\theta_{[j]}\partial\theta_{[l]}\partial\theta_{[m]}}=O_{p}(1)$, and $\frac{\partial^{4}Q_{n}(\theta_{1})}{\partial\theta_{[j]}\partial\theta_{[l]}\partial\theta_{[m]}\partial\theta_{[s]}}=O_{p}(1)$. Since we have shown $\hat{\theta}_{1}-\theta_{1}=O_{p}(n^{-1/2})$, it follows that $R_{n}^{*}=O_{p}(n^{-1})$. Note that this also justifies the expansion \eqref{Eq_feasible2_Expansion} by setting $\theta_{1}=\theta_{2}=\theta_{0}$ and acknowledging that $g_{n}(\theta_{0})=O_{p}(n^{-1/2})$ under correct specification.

By multiplying $\sqrt{n}$ on both sides \eqref{misexp} can now be written as
\begin{align}
\label{gmm2p} \sqrt{n}(\hat{\theta}_{2}-\theta_{2}) =& -(G'\Omega_{1}^{-1}G)^{-1}(I_{k}+o_{p}(1))G_{n}[\Omega_{n}(\theta_{1})]^{-1}\sqrt{n}g_{n}(\theta_{2})\\
\label{gmm2p1}& + D^{*}\sqrt{n}(\hat{\theta}_{1}-\theta_{1})+o_{p}(1).
\end{align}
Take \eqref{gmm2p}. By Lemma \ref{lemma_convergence} and the population FOC $G'\Omega_{1}^{-1}g_{2}=0$,
\begin{equation}
\label{foc2} G_{n}[\Omega_{n}(\theta_{1})]^{-1}g_{n}(\theta_{2}) =G'\Omega_{1}^{-1}g_{n}(\theta_{2}) + G_{n}'\Omega_{1}^{-1}g_{2} - G'\Omega_{1}^{-1}\Omega_{n}(\theta_{1})\Omega_{1}^{-1}g_{2} + O_{p}(n^{-1}).
\end{equation}
Let
\begin{equation*}
m_{2i} = G'\Omega_{1}^{-1}g(X_{i},\theta_{2}) + G(X_{i})'\Omega_{1}^{-1}g_{2} - G'\Omega_{1}^{-1}g(X_{i},\theta_{1})g(X_{i},\theta_{1})'\Omega_{1}^{-1}g_{2}.
\end{equation*}
Since $Em_{2i}=0$, by CLT under Assumption 1(v)
\begin{equation}
\frac{1}{\sqrt{n}}\sum_{i=1}^{n}\left(\begin{array}{c}
m_{2i}\\m_{1i}
\end{array}\right)\xrightarrow{d}N\left(0,\left(\begin{array}{cc}
Em_{2i}m_{2i}' & Em_{2i}m_{1i}'\\
Em_{1i}m_{2i}' & Em_{1i}m_{1i}'
\end{array}\right)\right)
\end{equation}
as $n\rightarrow\infty$. 
Replacing \eqref{foc2} and \eqref{gmm1} into \eqref{gmm2p} and \eqref{gmm2p1}, respectively, to obtain
\begin{align}
\notag \sqrt{n}(\hat{\theta}_{2}-\theta_{2}) =&-\left[\begin{array}{cc}
(G'\Omega_{1}^{-1}G)^{-1} & D^{*} (G'W^{-1}G)^{-1}
\end{array}\right]\frac{1}{\sqrt{n}}\sum_{i=1}^{n}\left(\begin{array}{c}
m_{2i}\\m_{1i}
\end{array}\right)+o_{p}(1)\\
\label{t2Op}\xrightarrow{d}&N(0,V_{2})
\end{align}
as $n\rightarrow\infty$ where $V_{2}$ is the asymptotic variance matrix of the two-step GMM under misspecification given by
\begin{align*}
V_{2} =& (G'\Omega_{1}^{-1}G)^{-1}Em_{2i}m_{2i}'(G'\Omega_{1}^{-1}G)^{-1} + D^{*}(G'W^{-1}G)^{-1}Em_{1i}m_{2i}'(G'\Omega_{1}^{-1}G)^{-1}\\
&+(G'\Omega_{1}^{-1}G)^{-1}Em_{2i}m_{1i}'(G'W^{-1}G)^{-1}D^{*'} + D^{*}(G'W^{-1}G)^{-1}Em_{1i}m_{1i}'(G'W^{-1}G)^{-1}D^{*'}.
\end{align*}
Since we have shown the asymptotic variance matrices of the GMM estimators robust to misspecification, the first part of the proof is complete. 

Note that $m_{1i} = G'W^{-1}g(X_{i},\theta_{0})$, $m_{2i} = G'\Omega_{0}^{-1}g(X_{i},\theta_{0})$, and $D^{*}=0$ under correct specification and $V_{1}$ and $V_{2}$ coincide to the conventional asymptotic variance matrices.

Next we show that $\widehat{V}_{dc}(\hat{\theta}_{1})\xrightarrow{p}V_{1}$ and $\widehat{V}_{dc}(\hat{\theta}_{2})\xrightarrow{p}V_{2}$. Since \eqref{gwg} and \eqref{gog} hold, by the CMT it suffices to show
\begin{align}
\label{m2m2}&\frac{1}{n}\sum_{i=1}^{n}m_{i}(\hat{\theta}_{2},\Omega_{n}(\hat{\theta}_{1}))m_{i}(\hat{\theta}_{2},\Omega_{n}(\hat{\theta}_{1}))'\xrightarrow{p}Em_{2i}m_{2i}',\\
\label{m1m2}&\frac{1}{n}\sum_{i=1}^{n}m_{i}(\hat{\theta}_{1},W_{n})m_{i}(\hat{\theta}_{2},\Omega_{n}(\hat{\theta}_{1}))'\xrightarrow{p}Em_{1i}m_{2i}',\\
\label{m1m1}&\frac{1}{n}\sum_{i=1}^{n}m_{i}(\hat{\theta}_{1},W_{n})m_{i}(\hat{\theta}_{1},W_{n})'\xrightarrow{p}Em_{1i}m_{1i}',\\
\label{Dcon}&\widehat{D}_{n}\xrightarrow{p}D^{*}.
\end{align}
\eqref{m2m2}-\eqref{m1m1} follow if we show
\begin{align}
\label{m22}m_{i}(\hat{\theta}_{2},\Omega_{n}(\hat{\theta}_{1})) = m_{2i} + O_{p}(n^{-1/2}),\\
\label{m11}m_{i}(\hat{\theta}_{1},W_{n}) = m_{1i} + O_{p}(n^{-1/2}),
\end{align}
Since by Lemma \ref{lemma_convergence}, \eqref{t1Op}, and \eqref{t2Op}, for $j=1,2$,
\begin{align*}
g_{n}(\hat{\theta}_{j}) =& g_{j} + (g_{n}(\theta_{j})-g_{j}) + G(\hat{\theta}_{j}-\theta_{j}) + (G_{n}-G)(\hat{\theta}_{j}-\theta_{j})\\
=&g_{j} + O_{p}(n^{-1/2}) 
\end{align*}
and 
\begin{equation*}
g(X_{i},\hat{\theta}_{j}) = g(X_{i},\theta_{j}) + G(X_{i})(\hat{\theta}_{j}-\theta_{j}) = g(X_{i},\theta_{j}) +O_{p}(n^{-1/2}).
\end{equation*}
Thus, \eqref{m22} and \eqref{m11} follow. To show \eqref{Dcon}, we use \eqref{gn212}, \eqref{Gn12}, \eqref{DOn12}, \eqref{iOnhat12} of Lemma \ref{lemma_convergence}.\qed

\newpage

\section*{Appendix B: Stochastic Expansion under Local Misspecification \label{Stochastic}}

In this section, we provide formal stochastic expansions of GMM estimators assuming that the moment condition evaluated at the true value is a local drifting sequence around zero:
\begin{equation}
E[g(X_{in},\theta_{0})]=\frac{\delta}{\sqrt{n}}%
\label{Eqn_local_moment_conditions}%
\end{equation}
for some nonzero $\delta\in\mathbb{R}^{q}$ that depends on $\theta_{0}.$ Note that the observations now form a triangular array $\{X_{in}:i=1,...,n,$ $n\in\mathbb{N}\}$ because \eqref{Eqn_local_moment_conditions} changes with $n$. Under \eqref{Eqn_local_moment_conditions} and regularity conditions, $\hat{\theta}_{1}$ and $\hat{\theta}_{2}$ are both consistent for $\theta_{0}$ and 
\begin{equation}
\sqrt{n}(g_{n}(\theta_{0})-E[g(X_{in},\theta_{0})]) = O_{p}(1).
\end{equation}
This setup is referred to as local misspecification in the literature, e.g., Newey (1985), Otsu (2011), Guggenberger (2012), Conley, Hansen, and Rossi (2012), Andrews, Gentzkow, and Shapiro (2017), Bonhomme and Weidner (2018), and Armstrong and Koles\'{a}r (2019). Since $g_{n}(\theta_{0})$ is still $O_{p}(n^{-1/2})$, the local misspecification and over-identification bias are essentially equivalent in stochastic orders.

Under correct specification \eqref{correct_spec}, Newey and Smith (2004) provide a thorough analysis on the higher-order bias and variance of the stochastic expansions of GMM and GEL estimators up to the order of $O_{p}(n^{-1})$ in the form of 
\begin{equation}
\label{Sto_Ex}
\sqrt{n}(\hat{\theta}-\theta_{0}) = \psi_{0} + \psi_{1}/\sqrt{n} + \psi_{2}/n + R_{n}
\end{equation}
where $\psi_{0}$, $\psi_{1}$, $\psi_{2}$ are all $O_{p}(1)$ and $R_{n} = O_{p}(n^{-3/2})$. Our analysis reveals additional higher-order terms of the order $O_{p}(n^{-1/2})$ in the RHS of \eqref{Sto_Ex} as well as the first order asymptotic bias in the point estimator. Specifically, we derive the exact expressions of $\psi_{0}$ and $\psi_{1}$ for the one-step and the two-step GMM to show that our double correction effectively estimates the (co)variance of some higher-order terms. Deriving the full expression of $\psi_{2}$ is not attempted because the derivation is not required to show the relationship between the double correction and the higher-order expansion.

Let $g(\theta) = E[g(X_{in},\theta)]$ and write $g=g(\theta_{0})$, $G = E[G(X_{in})]$, and $W = E[W(X_{in})]$. Define 
\begin{align*}
\widetilde{g}(\theta) =& \sum_{i=1}^{n}\left(g(X_{in},\theta)-g(\theta)\right)/\sqrt{n},\\
\widetilde{G} =& \sum_{i=1}^{n}(G(X_{in})-G)/\sqrt{n},\\
\widetilde{W} =& \sum_{i=1}^{n}(W(X_{in})-W)/\sqrt{n},\\
\widetilde{\Omega}(\theta) =& \sum_{i=1}^{n}(g(X_{in},\theta)g(X_{in},\theta)'-\Omega(\theta))/\sqrt{n},
\end{align*}
and write $\widetilde{g} = \widetilde{g}(\theta_{0})$ and $\widetilde{\Omega} = \widetilde{\Omega}(\theta_{0})$. First we consider the one-step GMM estimator. 

\begin{theorem}\label{thm:onestep_stochastic}
	Suppose that \eqref{Eqn_local_moment_conditions} holds, $\widetilde{g}=O_{p}(1)$, $\widetilde{G}=O_{p}(1)$, $\widetilde{W}=O_{p}(1)$, the FOC of the one-step GMM holds with probability approaching one (w.p.a. 1), $G$ is full column rank, and $W>0$. In addition, suppose that the second moment of $g(X_{in},\theta_{0})$, $G(X_{in})$, and $W(X_{in})$ exist and are finite. Then, the one-step GMM estimator has the following expansion w.p.a.1.
	\begin{equation}
	\label{thm:exp1}
	\sqrt{n}(\hat{\theta}_{1}-\theta_{0}) = \eta_{W} + \widetilde{\psi}_{W,0} + \frac{1}{\sqrt{n}}\left(\widetilde{\psi}_{W,1} + \widetilde{q}_{W} + \widetilde{B}_{W}\left(\eta_{W}+\widetilde{\psi}_{W,0}\right)\right) + O_{p}\left(\frac{1}{n}\right)
	\end{equation}
	where 
	\begin{align}
	\eta_{W} &= -(G'W^{-1}G)^{-1}G'W^{-1}\delta,\\
	\widetilde{\psi}_{W,0} &= -(G'W^{-1}G)^{-1}G'W^{-1}\widetilde{g},\\
	\widetilde{\psi}_{W,1} &= -(G'W^{-1}G)^{-1}\left(\widetilde{G}'W^{-1}\delta - G'W^{-1}\widetilde{W}W^{-1}\delta\right),\\
	\widetilde{q}_{W} &= -(G'W^{-1}G)^{-1}\left(\widetilde{G}W^{-1}\widetilde{g} - G^{\prime}W^{-1}\widetilde{W}W^{-1}\widetilde{g}\right)\\
	\widetilde{B}_{W} &= -(G'W^{-1}G)^{-1}\left(\widetilde{G}'W^{-1}G - G'W^{-1}\widetilde{W}W^{-1}G + G'W^{-1}\widetilde{G}\right).
	\end{align}
\end{theorem}

The first term in the expansion of Theorem \ref{thm:onestep_stochastic}, $\eta_{W}$, is the constant bias due to local misspecification. It shifts the mean of the distribution but does not alter the first order variance. This bias cannot be consistently estimated from the data in general. Conley, Hansen, and Rossi (2012) impose a prior distribution on $\delta$ and Armstrong and Koles\'{a}r (2019) set a pre-specified bound on $\delta$ to construct bias-corrected confidence intervals.

The conventional variance estimator estimates the variance of $\widetilde{\psi}_{W,0}$. The doubly corrected variance estimator \eqref{Vdc1} estimates the variance of $\widetilde{\psi}_{W,0} + \widetilde{\psi}_{W,1}/\sqrt{n}$. Since the variance of $\widetilde{\psi}_{W,1}$ increases with $\delta$, the double correction would be more effective with a larger $\delta$.

On the other hand, the double correction omits some terms of the same order, $(\widetilde{q}_{W} +  \widetilde{B}_{W}(\eta_{W}+\widetilde{\psi}_{W,0}))/\sqrt{n}$. We discuss these terms one by one. First, the variance of $(\widetilde{q}_{W}+\widetilde{B}_{W}\widetilde{\psi}_{W,0})/\sqrt{n}$ does not increase with $\delta$ because $\widetilde{g}$ is a centered process. Second, $\text{Var}(\widetilde{B}_{W}\eta_{W}/\sqrt{n})$ increases with $\delta$ but its sample analogue is zero because the sample analogue of $\eta_{W}$ is zero (FOC). So the (co)variance of this term is effectively estimated by zero in the double correction.

An estimator of the higher-order variance up to $O(n^{-1})$  would include the variance of $(\widetilde{q}_{W} + \widetilde{B}_{W}(\eta_{W}+\widetilde{\psi}_{W,0}))/\sqrt{n}$ and the covariance between $\widetilde{\psi}_{W,0}$ and $(\widetilde{q}_{W} + \widetilde{B}_{W}(\eta_{W}+\widetilde{\psi}_{W,0}))/\sqrt{n} + \widetilde{\psi}_{W,2}/n$ where $\widetilde{\psi}_{W,2}$ is the higher-order term corresponding $\psi_{2}$ in \eqref{Sto_Ex}. Since these terms are not considered, the doubly corrected variance estimator is not an estimator of the higher-order variance of Newey and Smith (2004). Although the higher-order variance up to $O(n^{-1})$ could be estimated by further expanding \eqref{thm:exp1} up to the remainder term of the order $O_{p}(n^{-3/2})$ the resulting estimator would include many terms to be estimated which would not be practical. Thus, the doubly corrected variance estimator can be viewed as a convenient alternative to the higher-order variance estimator.

Next we consider the two-step GMM estimator.
\begin{theorem} \label{thm:stochastic_expansion} Suppose that the assumptions of Theorem \ref{thm:onestep_stochastic} hold. In addition, $\widetilde{\Omega}=O_{p}(1)$, the FOC of the two-step GMM holds w.p.a.1, $\Omega>0$, and the fourth moment of $g(X_{in},\theta_{0})$ and $G(X_{in})$ exist and are finite. Then, the two-step GMM estimator has the following expansion w.p.a.1.
	\begin{align*}
	&\sqrt{n}(\hat{\theta}_{2}-\theta_{0})\\
	=& \left[\eta_{\Omega} + \frac{1}{\sqrt{n}}\left(\left(D + H_{\eta_{\Omega}}\right)\eta_{W}\right)\right]\\
	& +\widetilde{\psi}_{\Omega,0}+\frac{1}{\sqrt{n}}\left(\widetilde{\psi}_{\Omega,1} + \left(D +\widetilde{C}+ H_{\eta_{\Omega}}+H_{\widetilde{\psi}_{\Omega,0}}\right)\widetilde{\psi}_{W,0}+ \widetilde{q}_{\Omega} + \widetilde{B}_{\Omega}\left(\eta_{\Omega}+\widetilde{\psi}_{\Omega,0}\right)+ \left(\widetilde{C}+H_{\widetilde{\psi}_{\Omega,0}}\right)\eta_{W} \right)\\
	&+\frac{1}{n}D\widetilde{\psi}_{W,1}+O_{p}\left(\frac{1}{n}\right)
	\end{align*}
	where
	\begin{align}
	\eta_{\Omega} &= -(G'\Omega^{-1}G)^{-1}G'\Omega^{-1}\delta,\\
	\widetilde{\psi}_{\Omega,0} &= -(G'\Omega^{-1}G)^{-1}G'\Omega^{-1}\widetilde{g},\\
	\widetilde{\psi}_{\Omega,1} &= -(G'\Omega^{-1}G)^{-1}\left(\widetilde{G}'\Omega^{-1}\delta- G'\Omega^{-1}\widetilde{\Omega}\Omega^{-1}\delta\right),\\
	\widetilde{q}_{\Omega} &= -(G'\Omega^{-1}G)^{-1}\left(\widetilde{G}'\Omega^{-1}\widetilde{g} - G'\Omega^{-1}\widetilde{\Omega}\Omega^{-1}\widetilde{g}\right),\\
	\widetilde{B}_{\Omega} &= -(G'\Omega^{-1}G)^{-1}\left(\widetilde{G}'\Omega^{-1}G - G'\Omega^{-1}\widetilde{\Omega}\Omega^{-1}G + G'\Omega^{-1}\widetilde{G}\right),\\
	D[.,j] &= (G'\Omega^{-1}G)^{-1}G'\Omega^{-1}\left.\frac{\partial\Omega(\theta)}{\partial\theta_{[j]}}\right|_{\theta=\theta_{0}}\Omega^{-1}\delta,\\
	\widetilde{C}[.,j] &= (G'\Omega^{-1}G)^{-1}G'\Omega^{-1}\left.\frac{\partial\Omega(\theta)}{\partial\theta_{[j]}}\right|_{\theta=\theta_{0}}\Omega^{-1}\widetilde{g},\\
	H_{v}[.,j] &= (G'\Omega^{-1}G)^{-1}G'\Omega^{-1}\left.\frac{\partial\Omega(\theta)}{\partial\theta_{[j]}}\right|_{\theta=\theta_{0}}\Omega^{-1}Gv.
	\end{align}
\end{theorem}

The terms in the square brackets in the expansion, $\eta_{\Omega} + (D+H_{\eta_{\Omega}})\eta_{W}/\sqrt{n}$ is the constant bias due to local misspecification which cannot be consistently estimated in general. It only shifts the mean of the distribution but does not affect the variance. 

The conventional, the Windmeijer correction, and the doubly corrected variance estimators estimate the variance of
\begin{align*}
\text{conventional: }&\widetilde{\psi}_{\Omega,0},\\
\text{Windmeijer: }& \widetilde{\psi}_{\Omega,0} +\frac{D\widetilde{\psi}_{W,0}}{\sqrt{n}},\\
\text{double correction: }& \widetilde{\psi}_{\Omega,0}  + \frac{\widetilde{\psi}_{\Omega,1} +D\widetilde{\psi}_{W,0}}{\sqrt{n}}+ \frac{D\widetilde{\psi}_{W,1}}{n},
\end{align*}
respectively. Similar to the one-step GMM expansion, the double correction estimates the variance of the higher-order terms that increase with $\delta$: $D$, $\widetilde{\psi}_{\Omega,1}$, and $\widetilde{\psi}_{W,1}$. In contrast, some (or all) of these terms are omitted in the Windmeijer correction (or the conventional variance estimator). This implies that those variance estimators tend to differ more with a larger $\delta$, and we show that including these terms in the double correction formula is important to get robustness to global (fixed) misspecification in Section \ref{Misspecification}. This is also supported by the simulation experiment in Section \ref{Section: Simulation}.

By a similar argument with the one-step GMM, both the doubly corrected and the Windmeijer corrected variance estimators are not the higher-order variance estimator of Newey and Smith (2004) because some higher-order terms up to the order $O_{p}(n^{-1})$, which will deliver higher-order variance, are omitted in both corrections.

\bigskip

\noindent\textbf{Remark (Edgeworth expansion)} The improved approximation to the finite sample variance of the GMM estimators by the double correction is different from the higher-order refinement via the Edgeworth expansion which expands the finite sample distribution function of the (standardized) test statistic. Using analytical expansions (Rothenberg, 1984; Hansen, 2006; Kundhi and Rilstone, 2013) or the bootstrap (Hall, 1992; Hall and Horowitz, 1996; Andrews, 2002; Lee, 2014, 2016), the resulting critical value gives a smaller error in the size of the test or in the coverage probability of the confidence interval.

\bigskip 

\noindent
\textbf{Proof of Theorem \ref{thm:onestep_stochastic}: } In what following the statements hold with probability approaching one. First note that 
\begin{align*}
G_{n}'W_{n}^{-1}G_{n} &= G'W^{-1}G+ (G_{n}-G)'W^{-1}G + G'W^{-1}(G_{n}-G) + G'(W_{n}^{-1}-W^{-1})G + O_{p}(n^{-1})\\
&=G'W^{-1}G + O_{p}(n^{-1/2}).
\end{align*}
By Lemma 1 with $q=1$,
\begin{align*}
(G_{n}'W_{n}^{-1}G_{n})^{-1} =& \left(G'W^{-1}G + \frac{1}{\sqrt{n}}\sqrt{n}(G_{n}'W_{n}^{-1}G_{n}-G'W^{-1}G)\right)^{-1}\\
=& (G'W^{-1}G)^{-1} - \frac{1}{\sqrt{n}}(G'W^{-1}G)^{-1}\sqrt{n}(G_{n}'W_{n}^{-1}G_{n}-G'W^{-1}G)(G'W^{-1}G)^{-1}\\
& + O_{p}(n^{-1})\\
=& (G'W^{-1}G)^{-1} + \frac{1}{\sqrt{n}}\widetilde{B}_{W}(G'W^{-1}G)^{-1}+ O_{p}(n^{-1}).
\end{align*}
In addition,
\begin{align*}
G_{n}'W_{n}^{-1}\sqrt{n}g_{n}(\theta_{0}) =& G_{n}'W_{n}^{-1}\sqrt{n}(g_{n}(\theta_{0})-g) + \sqrt{n}(G_{n}-G)'W_{n}^{-1}g + G'\sqrt{n}(W_{n}^{-1}-W^{-1})g + G'W^{-1}\delta\\
=& G'W^{-1}\widetilde{g} + \widetilde{G}'W^{-1}g - G'W^{-1}\widetilde{W}W^{-1}g + \widetilde{G}'W^{-1}\widetilde{g}/\sqrt{n} - G'W^{-1}\widetilde{W}W^{-1}\widetilde{g}/\sqrt{n}\\
& + G'W^{-1}\delta + O_{p}(n^{-1}).
\end{align*}

Since the FOC of the one-step GMM holds regardless of misspecification of the moment condition, we use the above expansions to obtain
\begin{align*}
&\sqrt{n}(\hat{\theta}_{1}-\theta_{0})\\
=&(G_{n}W_{n}^{-1}G_{n})^{-1}G_{n}W_{n}^{-1}\sqrt{n}g_{n}(\theta_{0})\\
=&-\left[  (G'W^{-1}G)^{-1} + \widetilde{B}_{W}(G'W^{-1}G)^{-1}/\sqrt{n}\right]\times \left[G'W^{-1}\widetilde{g}+ \widetilde{G}'W^{-1}g - G'W^{-1}\widetilde{W}W^{-1}g \right.\\
& \left.+ \widetilde{G}'W^{-1}\widetilde{g}/\sqrt{n} - G'W^{-1}\widetilde{W}W^{-1}\widetilde{g}/\sqrt{n}+ G'W^{-1}\delta\right]+ O_{p}(n^{-1})\\
=& \eta_{W} + \widetilde{B}_{W}\eta_{W}/\sqrt{n} -(G'W^{-1}G)^{-1}G^{\prime}W^{-1}\widetilde{g} -(G'W^{-1}G)^{-1}\left(\widetilde{G}'W^{-1}g - G'W^{-1}\widetilde{W}W^{-1}g\right)\\
&- (G'W^{-1}G)^{-1}\left(\widetilde{G}W^{-1}\widetilde{g} - G^{\prime}W^{-1}\widetilde{W}W^{-1}\widetilde{g}\right)/\sqrt{n}-\widetilde{B}_{W}(G'W^{-1}G)^{-1}G'W^{-1}\widetilde{g}/\sqrt{n} + O_{p}(n^{-1})\\
=& \eta_{W} + \widetilde{\psi}_{W,0} + (\widetilde{\psi}_{W,1} + \widetilde{q}_{W} + \widetilde{B}_{W}(\widetilde{\psi}_{W,0}+\eta_{W}))/\sqrt{n} + O_{p}(n^{-1}).
\end{align*}
\qed

\bigskip

\noindent
\textbf{Proof of Theorem \ref{thm:stochastic_expansion}:} In what following the statements hold with probability approaching one. We first prove some useful expansions. Note that from Theorem 1, $\sqrt{n}(\hat{\theta}_{1}-\theta_{0})=O_{p}(1)$. By a similar argument with the proof of \eqref{Onhat12}, we can show
\begin{equation}
\label{Omegan1}
\Omega_{n}(\hat{\theta}_{1}) = \Omega + O_{p}(n^{-1/2}).
\end{equation}
By Lemma \ref{lemma_matrix} with $q=0$, 
\begin{align}
\label{O11} [\Omega_{n}(\hat{\theta}_{1})]^{-1} &= \Omega^{-1} + O_{p}(n^{-1/2}),\\
\label{O01} [\Omega_{n}(\theta_{0})]^{-1} &= \Omega^{-1} + O_{p}(n^{-1/2}).
\end{align}
Using \eqref{O01} and applying Lemma \ref{lemma_matrix} with $q=1$ to \eqref{Omegan1},
\begin{align}
[\Omega_{n}(\hat{\theta}_{1})]^{-1} &= [\Omega_{n}(\theta_{0})]^{-1} - \frac{1}{\sqrt{n}}[\Omega_{n}(\theta_{0})]^{-1}\sum_{j}\frac{\partial\Omega(\theta)}{\partial\theta_{[j]}}\sqrt{n}(\hat{\theta}_{1[j]}-\theta_{0[j]})[\Omega_{n}(\theta_{0})]^{-1} + O_{p}(n^{-1})\\
\label{O11_O01} & =  [\Omega_{n}(\theta_{0})]^{-1} - \frac{1}{\sqrt{n}}\Omega^{-1}\sum_{j}\frac{\partial\Omega(\theta)}{\partial\theta_{[j]}}\sqrt{n}(\hat{\theta}_{1[j]}-\theta_{0[j]})\Omega^{-1} + O_{p}(n^{-1}).
\end{align}
Using a similar argument with the proof of Theorem \ref{thm:onestep_stochastic} and Lemma \ref{lemma_matrix} with $q=1$, we obtain
\begin{align}
\label{GOG1}(G_{n}'[\Omega_{n}(\theta_{0})]^{-1}G_{n})^{-1} =& (G'\Omega^{-1}G)^{-1} + \frac{1}{\sqrt{n}}B_{\Omega}(G'\Omega^{-1}G)^{-1}+ O_{p}(n^{-1}),\\
\label{GOg} G_{n}'[\Omega_{n}(\theta_{0})]^{-1}\sqrt{n}g_{n}(\theta_{0}) =& G'\Omega^{-1}\delta +G'\Omega^{-1}\widetilde{g} + \widetilde{G}'\Omega^{-1}g - G'\Omega^{-1}\widetilde{\Omega}\Omega^{-1}g\\
& +  \widetilde{G}\Omega^{-1}\widetilde{g}/\sqrt{n} - G'\Omega^{-1}\widetilde{\Omega}\Omega^{-1}\widetilde{g}/\sqrt{n} + O_{p}(n^{-1}).
\end{align}

Note that the assumption of the theorem implies that $\hat{\theta}_{2}-\theta_{0} = O_{p}(n^{-1/2})$. Using \eqref{O11_O01}, the first-order Taylor expansion of the FOC of $\hat{\theta}_{2}$ around $\theta_0$ can be written as
\begin{align}
0 =& G_{n}^{\prime}[\Omega_n(\hat{\theta}_1)]^{-1}g_{n}(\theta_0) + G_{n}^{\prime}[\Omega_n(\hat{\theta}_1)]^{-1} G_{n} (\hat{\theta}_{2} - \theta_{0})\\
=& G_{n}^{\prime}[\Omega_n(\theta_0)]^{-1}g_{n}(\theta_0) + G_{n}^{\prime}[\Omega_n(\theta_0)]^{-1} G_{n} (\hat{\theta}_{2} - \theta_{0})\\
&+ G_{n}'\left([\Omega_n(\hat{\theta}_1)]^{-1}-[\Omega_n(\hat{\theta}_0)]^{-1}\right)\left(g_{n}(\theta_{0}) + G_{n}(\hat{\theta}_{2}-\theta_{0})\right)\\ 
=& G_{n}^{\prime}[\Omega_n(\theta_0)]^{-1}g_{n}(\theta_0) + G_{n}^{\prime}[\Omega_n(\theta_0)]^{-1} G_{n} (\hat{\theta}_{2} - \theta_{0})\\
&- \frac{1}{\sqrt{n}}G'\Omega^{-1}\sum_{j}\frac{\partial\Omega(\theta)}{\partial\theta_{[j]}}\sqrt{n}(\hat{\theta}_{1[j]}-\theta_{0[j]})\Omega^{-1}\left(g_{n}(\theta_{0}) + G(\hat{\theta}_{2}-\theta_{0})\right) + O_{p}(n^{-3/2}).
\end{align}
By arranging terms, multiplying $\sqrt{n}$, using \eqref{GOG1}-\eqref{GOg},
\begin{align*}
\sqrt{n}(\hat{\theta}_{2}-\theta_{0}) =& -\left(G_{n}^{\prime}[\Omega_n(\theta_0)]^{-1} G_{n} \right)^{-1}G_{n}^{\prime}[\Omega_n(\theta_0)]^{-1}\sqrt{n}g_{n}(\theta_{0})\\
&+\left(G_{n}^{\prime}[\Omega_n(\theta_0)]^{-1} G_{n} \right)^{-1}\\
&\times \frac{1}{\sqrt{n}}G'\Omega^{-1}\sum_{j}\frac{\partial\Omega(\theta)}{\partial\theta_{[j]}}\sqrt{n}(\hat{\theta}_{1[j]}-\theta_{0[j]})\Omega^{-1}\left(\sqrt{n}g_{n}(\theta_{0}) + G\sqrt{n}(\hat{\theta}_{2}-\theta_{0})\right) + O_{p}(n^{-1}).
\end{align*}
Note that the second term in the RHS is $O_{p}(n^{-1/2})$. First,
\begin{align}
&-\left(G_{n}^{\prime}[\Omega_n(\theta_0)]^{-1} G_{n} \right)^{-1}G_{n}^{\prime}[\Omega_n(\theta_0)]^{-1}\sqrt{n}g_{n}(\theta_{0})\\
=&-\left((G'\Omega^{-1}G)^{-1} + \frac{1}{\sqrt{n}}\widetilde{B}_{\Omega}(G'\Omega^{-1}G)^{-1}\right)\\
&\times \left(G'\Omega^{-1}\delta +G'\Omega^{-1}\widetilde{g} + \widetilde{G}'\Omega^{-1}g - G'\Omega^{-1}\widetilde{\Omega}\Omega^{-1}g+  \widetilde{G}\Omega^{-1}\widetilde{g}/\sqrt{n} - G'\Omega^{-1}\widetilde{\Omega}\Omega^{-1}\widetilde{g}/\sqrt{n} \right)\\
&+O_{p}(n^{-1})\\
\label{exp1} =& \eta_{\Omega} + \frac{1}{\sqrt{n}}\widetilde{B}_{\Omega}\eta_{\Omega}+ \widetilde{\psi}_{\Omega,0} + \frac{1}{\sqrt{n}}\left(\widetilde{\psi}_{\Omega,1} + \widetilde{q}_{\Omega} + \widetilde{B}_{\Omega}\widetilde{\psi}_{\Omega,0}\right) + O_{p}(n^{-1}).
\end{align}
Thus, we have $\sqrt{n}(\hat{\theta}_{2}-\theta_{0}) = \eta_{\Omega} + \widetilde{\psi}_{\Omega,0} + O_{p}(n^{-1/2})$. Next, using $\sqrt{n}(\hat{\theta}_{1}-\theta_{0}) = \eta_{W} + \widetilde{\psi}_{W,0} + O_{p}(n^{-1/2})$,
\begin{align}
\label{exp2} &\left(G_{n}^{\prime}[\Omega_n(\theta_0)]^{-1} G_{n} \right)^{-1}\frac{1}{\sqrt{n}}G'\Omega^{-1}\sum_{j}\frac{\partial\Omega(\theta)}{\partial\theta_{[j]}}\sqrt{n}(\hat{\theta}_{1[j]}-\theta_{0[j]})\Omega^{-1}\left(\sqrt{n}g_{n}(\theta_{0}) + G\sqrt{n}(\hat{\theta}_{2}-\theta_{0})\right)\\
=&\left(G^{\prime}\Omega^{-1} G \right)^{-1}\frac{1}{\sqrt{n}}G'\Omega^{-1}\sum_{j}\frac{\partial\Omega(\theta)}{\partial\theta_{[j]}}(\eta_{W[j]}+\widetilde{\psi}_{W,0[j]})\Omega^{-1}\left(\widetilde{g} + \delta + G(\eta_{\Omega}+\widetilde{\psi}_{\Omega,0})\right) + O_{p}(n^{-1})\\
=&\left(G^{\prime}\Omega^{-1} G \right)^{-1}\frac{1}{\sqrt{n}}G'\Omega^{-1}\sum_{j}\frac{\partial\Omega(\theta)}{\partial\theta_{[j]}}\eta_{W[j]}\Omega^{-1}\left(\widetilde{g} + \delta + G(\eta_{\Omega}+\widetilde{\psi}_{\Omega,0})\right) \\
&+\left(G^{\prime}\Omega^{-1} G \right)^{-1}\frac{1}{\sqrt{n}}G'\Omega^{-1}\sum_{j}\frac{\partial\Omega(\theta)}{\partial\theta_{[j]}}\widetilde{\psi}_{W,0[j]}\Omega^{-1}\left(\widetilde{g} + \delta + G(\eta_{\Omega}+\widetilde{\psi}_{\Omega,0})\right) + O_{p}(n^{-1})\\
=&\frac{1}{\sqrt{n}}\left\{\left(D + H_{\eta_{\Omega}}\right)\eta_{W} + \left(\widetilde{C}+H_{\widetilde{\psi}_{\Omega,0}}\right)\left(\eta_{W} + \widetilde{\psi}_{W,0}\right)  + \left(D + H_{\eta_{\Omega}}\right)\widetilde{\psi}_{W,0}\right\}+ O_{p}(n^{-1}).
\end{align}
Combining this with \eqref{exp1}, 
\begin{align}
&\sqrt{n}(\hat{\theta}_{2}-\theta_{0})\\
=& \eta_{\Omega} + \frac{1}{\sqrt{n}}\left(\left(D + H_{\eta_{\Omega}}\right)\eta_{W}\right)+ \widetilde{\psi}_{\Omega,0}\\
&+\frac{1}{\sqrt{n}}\left(\widetilde{\psi}_{\Omega,1} + \left(D+\widetilde{C} + H_{\eta_{\Omega}}+H_{\widetilde{\psi}_{\Omega,0}}\right)\widetilde{\psi}_{W,0}+ \widetilde{q}_{\Omega} + \widetilde{B}_{\Omega}\left(\eta_{\Omega}+\widetilde{\psi}_{\Omega,0}\right)+ \left(\widetilde{C}+H_{\widetilde{\psi}_{\Omega,0}}\right)\eta_{W}   \right)\\
&+O_{p}(n^{-1}).
\end{align}
Notice that $D\widetilde{\psi}_{W,1}/n$ in the $O_{p}(n^{-1})$ remainder term can be obtained if we plug $\sqrt{n}(\hat{\theta}_{1}-\theta_{0}) = \eta_{W} + \widetilde{\psi}_{W,0} + \widetilde{\psi}_{W,1}/\sqrt{n} + O_{p}(n^{-1/2})$ into \eqref{exp2}. This proves the theorem. \qed

\end{document}